\documentstyle[11pt,aaspp4,psfig,rotating]{article}

\lefthead{Anderson et al.}
\righthead{SDSS/ROSAT QSOs}

\begin{document}

\title{A Large, Uniform Sample of X-ray Emitting AGN:
Selection Approach and an Initial Catalog from the ROSAT All-Sky and Sloan
Digital Sky Surveys}
{
\author{Scott F. Anderson\altaffilmark{\ref{UW}}, 
Wolfgang Voges\altaffilmark{\ref{MPE}}, 
Bruce Margon\altaffilmark{\ref{STScI}}, 
Joachim Tr\"umper\altaffilmark{\ref{MPE}},
Marcel A. Ag\"ueros\altaffilmark{\ref{UW}},
Thomas Boller\altaffilmark{\ref{MPE}},
Matthew J. Collinge\altaffilmark{\ref{Princeton}},
L. Homer\altaffilmark{\ref{UW}},
Gregory Stinson\altaffilmark{\ref{UW}},
Michael~A.~Strauss\altaffilmark{\ref{Princeton}},
James~Annis\altaffilmark{\ref{FNAL}},
Percy Gomez\altaffilmark{\ref{CMU}},
Patrick B. Hall\altaffilmark{\ref{Princeton},\ref{CatChile}},
Robert C.~Nichol\altaffilmark{\ref{CMU}},
Gordon~T.~Richards\altaffilmark{\ref{Princeton}},
Donald~P.~Schneider\altaffilmark{\ref{PennState}},
Daniel~E.~Vanden~Berk\altaffilmark{\ref{Pitt}},
Xiaohui Fan\altaffilmark{\ref{UAz}},
\v Zeljko~Ivezi\'c\altaffilmark{\ref{Princeton}},
Jeffrey~A.~Munn\altaffilmark{\ref{USNOAZ}},
Heidi Jo Newberg\altaffilmark{\ref{RPI}},
Michael W. Richmond\altaffilmark{\ref{RochIT}},
David H. Weinberg\altaffilmark{\ref{OhioState}},
Brian~Yanny\altaffilmark{\ref{FNAL}},
Neta~A.~Bahcall\altaffilmark{\ref{Princeton}},
J.~Brinkmann\altaffilmark{\ref{APO}},
Masataka~Fukugita\altaffilmark{\ref{CosJapan}},
Donald~G.~York\altaffilmark{\ref{Chicago}}
}

\noindent
email addresses: anderson@astro.washington.edu, wvoges@mpe.mpg.de, margon@stsci.edu

\newcounter{address}
\setcounter{address}{1}
\altaffiltext{\theaddress}{University of Washington, Department of
   Astronomy, Box 351580, Seattle, WA 98195.
\label{UW}}
\addtocounter{address}{1}
\altaffiltext{\theaddress}{Max Planck-Institute f\"ur extraterrestrische
Physik, Geissenbachstr. 1, D-85741 Garching, Germany.
\label{MPE}}
\addtocounter{address}{1}
\altaffiltext{\theaddress}{Space Science Telescope Institute, 3700 San
Martin Drive, Baltimore, MD, 21218.
\label{STScI}}
\addtocounter{address}{1}
\altaffiltext{\theaddress}{Princeton University Observatory, Princeton,
   NJ 08544.
\label{Princeton}}
\addtocounter{address}{1}
\altaffiltext{\theaddress}{Fermi National Accelerator Laboratory, P.O. Box 500,
   Batavia, IL 60510.
\label{FNAL}}
\addtocounter{address}{1}
\altaffiltext{\theaddress}{Dept. of Physics, Carnegie Mellon University,
     5000~Forbes Ave., Pittsburgh, PA~15232.
\label{CMU}}
\addtocounter{address}{1}
\altaffiltext{\theaddress}{Departamento de Astronom\'{\i}a y Astrof\'{\i}sica, 
        Facultad de F\'{\i}sica, Pontificia Universidad Cat\'{o}lica de Chile, 
        Casilla 306, Santiago 22, Chile
\label{CatChile}}
\addtocounter{address}{1}
\altaffiltext{\theaddress}{Department of Astronomy and Astrophysics, The
   Pennsylvania State University, University Park, PA 16802.
\label{PennState}}
\addtocounter{address}{1}
\altaffiltext{\theaddress}{Dept. of Physics and Astronomy, University of
     Pittsburgh, Pittsburgh, PA~15260.
\label{Pitt}}
\addtocounter{address}{1}
\altaffiltext{\theaddress}{Steward Observatory, The University of Arizona,
        Tucson, AZ 85732.
\label{UAz}}
\addtocounter{address}{1}
\altaffiltext{\theaddress}{US Naval Observatory, Flagstaff Station,
P.O. Box 1149, Flagstaff, AZ 86002-1149.
\label{USNOAZ}}
\addtocounter{address}{1}
\altaffiltext{\theaddress}{Dept. of Physics, Applied Physics and
Astronomy, Rensselaer Polytechnic Institute, Troy NY 12180.
\label{RPI}}
\addtocounter{address}{1}
\altaffiltext{\theaddress}{Department of Physics, 
Rochester Institute of Technology, Rochester, NY 14623-5603.
\label{RochIT}}
\addtocounter{address}{1}
\altaffiltext{\theaddress}{Department of Astronomy, 
Ohio State University, Columbus, OH 43210.
\label{OhioState}}
\addtocounter{address}{1}
\altaffiltext{\theaddress}{Apache Point Observatory, P.O. Box 59,
Sunspot, NM 88349-0059.
\label{APO}}
\addtocounter{address}{1}
\altaffiltext{\theaddress}{Institute for Cosmic Ray Research, University
of Tokyo, Midori, Tanashi, Tokyo 188-8588, Japan
\label{CosJapan}}
\addtocounter{address}{1}
\altaffiltext{\theaddress}{Astronomy and Astrophysics Center, University of
   Chicago, 5640 South Ellis Avenue, Chicago, IL 60637.
\label{Chicago}}

\begin{abstract} Many open questions in X-ray astronomy are
limited by the relatively small number of objects in uniform
optically-identified and observed samples, especially when rare subclasses
are considered, or when subsets are isolated to search for evolution or
correlations between wavebands. We describe initial results of a new
program aimed to ultimately yield $\sim$10$^4$ fully characterized X-ray
source identifications---a sample about an order of magnitude larger than
earlier efforts. The technique is detailed, and employs X-ray data from
the ROSAT All-Sky Survey (RASS), and optical imaging and spectroscopic
follow-up from the Sloan Digital Sky Survey (SDSS); these two surveys
prove to be serendipitously very well matched in sensitivity. As part of
the SDSS software pipelines, optical objects in the SDSS photometric
catalogs are automatically positionally cross-correlated with RASS X-ray
sources. Then priorities for follow-on SDSS optical spectra of candidate
counterparts are automatically assigned using an algorithm based on the
known ratios of $f_{x}/f_{opt}$ for various classes of X-ray emitters at
typical RASS fluxes of $\sim10^{-13}$~erg~cm$^{-2}$~s$^{-1}$.  SDSS
photometric parameters for optical morphology, magnitude, colors, plus
FIRST radio information, serve as proxies for object class.

Initial application of this approach to RASS/SDSS data from 1400~deg$^2$
of sky provides a catalog of more than 1200 spectroscopically confirmed
quasars and other AGN that are probable RASS identifications. Most of
these are new identifications, and only a few percent of the AGN
counterparts are likely to be random superpositions. The magnitude and
redshift ranges of the counterparts are very broad, extending over
$15<m<21$ and $0.03<z<3.6$, respectively. Although most identifications
are quasars and Seyfert 1s, a variety of other AGN subclasses are also
sampled. Substantial numbers of rare AGN types are found, including more
than 130 narrow-line Seyfert~1s and 45 BL Lac candidates. These early
results already provide a very sizeable set of source identifications,
demonstrate utility of the sample in multi-waveband investigations, and
show the capability of the joint RASS/SDSS approach to efficiently proceed
towards the largest homogeneously selected/observed sample of X-ray
emitting quasars and other kinds of AGN.

\end{abstract}

\section{Introduction}

Although extrasolar X-ray sources were first observed four decades ago
(Giacconi et al. 1962), the first all-sky {\it imaging} X-ray survey has only 
recently been achieved. The ROSAT All-Sky Survey (hereafter RASS;
Voges et al. 1999, 2000) covers the entire 
celestial sphere in the $0.1-2.4$~keV range with the Position
Sensitive Proportional Counter (PSPC; Pfeffermann et al. 1988) to a typical limiting 
sensitivity of $\sim 10^{-13}$~erg~cm$^{-2}$~s$^{-1}$, although due to 
the scanning protocol, the exposure times and thus sensitivity limits vary 
markedly from the ecliptic pole to equator. Depending on the level of 
statistical significance discussed, $10^4$--$10^5$ X-ray sources are contained 
in the RASS Bright and Faint Source Catalogs, with more than 124,000 
sources in the catalog versions considered here. Positional 
accuracies derivable for the sources vary with count rate, but moderately 
well-exposed sources typically have a positional 
uncertainty of $\sim10-30''$. 
Many important inferences can and have been made from RASS data based directly 
on the X-ray observations alone, with minimal need for new correlative 
observations at other wavelengths. For example, the RASS point source 
catalogs provide an exceptional measure of a portion
of the X-ray log~N/log~S diagram (e.g., Voges et al. 1999).

However, the scale of the effort involved in identifying a large fraction of 
the abundant RASS sources, especially those in the Faint Source
Catalog, poses an unusual analysis problem. From ROSAT and also previous 
generations of X-ray observatories, especially the extensive observations of 
the {\it Einstein Observatory} (e.g., Gioia et al. 1984), such X-ray source 
counterparts are known to 
include a highly heterogeneous mix of objects, ranging from nearby 
M~dwarfs to distant quasars. In many cases, the X-ray data taken alone 
cannot unambiguously determine whether an X-ray source is Galactic or 
extragalactic, much less finer distinctions about its nature.
It has thus long been realized that identification of optical 
counterparts is an essential companion study to large X-ray surveys.

The most complete optical identification effort attempted on {\it Einstein}
X-ray data was that of the Extended Medium
Sensitivity Survey (EMSS; e.g., Stocke et al. 1991).
The comparison with RASS is not inappropriate, as
although {\it Einstein} surveyed $<10\%$ of the celestial sphere, the flux
levels and positional accuracy of source determinations from the 
{\it Einstein} Imaging Proportional Counter detector were not greatly 
dissimilar to the ROSAT PSPC (though, for example, the latter provides better
positional information). Optical identification of sources in the EMSS have been 
made in an exceptionally ambitious, decade-long program 
(e.g., Stocke et al. 1991 and references therein), with confirming optical 
spectra taken one at a time, using a heterogeneous collection of telescopes
and detectors, the only plausible facilities available at that time.
The EMSS yielded an impressive $\sim800$ optical counterparts, including 
normal stellar coronae, pre-main-sequence stars, interacting binaries, 
supernova remnants, nearby galaxies, clusters of galaxies, and, especially, 
active galactic nuclei (AGN) of a broad range of luminosities. Quasars/AGN
are the predominant class in the EMSS, accounting for more than $400$ of the 
{\it Einstein} identifications.

In a similar effort devoted to ROSAT data, there have been 
successful optical identification programs conducted, or still in progress,
on bright subsets of the RASS, or complete subsets over small regions
of sky, by a large collaboration (Fischer et al. 1998, Zickgraf et
al. 1998, Schwope et al. 2000). The goal is again $\sim$10$^3$ optical
identifications. In addition, a remarkably resourceful effort using the Hamburg
objective prism plates (Bade et al. 1998a, Zickgraf et al. 2003) has resulted in the tabulation
of candidate identifications for a significant number of the brightest 
RASS sources. Typically AGN are inferred as the correct identification if an
object with $B<18.5$ has a blue continuum color over the 2000~\AA~of plate
coverage; often emission lines are not detectable. However even for this
brighter subset, accurate redshifts, magnitudes, optical continuum slopes, and
emission line identifications, intensities, and equivalent widths are
largely unavailable.

Although the $\sim$10$^3$ total identifications from $\sim10^3$ deg$^2$ of
sky provided by the EMSS (and similar later ROSAT surveys) is an extremely
respectable number, there remain scientific problems where the size of an
identified {\it subclass} of X-ray emitter of interest is uncomfortably
small. Notable examples include studies of the more rare subclasses of AGN, 
such as narrow-line Seyfert~1s (hereafter, NLS1s) and BL~Lacs. For example, 
among the large sample of more than 
400 AGN from the EMSS, only about $\sim$40 BL~Lacs were found; and yet, even 
this small subsample suggested remarkable ``negative evolution" 
(e.g., Morris et al. 1991) that demands further study.
If such rare subclasses of AGN evolve in 
their properties with redshift, and/or exhibit complex dependencies of 
X-ray luminosity with optical and/or radio luminosity, as more typical 
quasars are known to do (e.g., Avni \& Tananbaum 1986; Wilkes et al. 1994;
Avni, Worrall, \& Morgan 1995), then an already small 
subsample must be further subdivided into yet smaller bins for 
analysis, leaving literally a handful of objects per bin. Another, less 
obvious, example is that although modest- to high-redshift quasars are not 
usually thought of as ``rare", X-ray selection at the {\it Einstein} and 
RASS flux levels strongly favors the discovery of nearby and low luminosity
AGN; in fact, the entire EMSS sample of $>400$ AGN includes only a small 
handful of X-ray emitting quasars with $z>2$, thus limiting its utility for 
studies of the quasar X-ray luminosity function at moderate- to high-redshift.

In order to effectively build and expand upon such earlier large scale 
counterpart identification programs, any new effort (at comparable X-ray 
depth) must survey substantially in excess of the $\sim10^3$ deg$^2$ covered by 
the EMSS, and must ultimately provide a sample numbering substantially in 
excess 
of $\sim10^3$ optical identifications, obtained in uniform 
fashion. Moreover, if many more than $10^3$ counterparts are to be newly found
on a reasonable timescale, more efficient optical identification techniques 
than previously available must be used.
The requisite large X-ray sample is now available via the RASS Bright and 
Faint Source Catalogs; the more efficient optical identification and analysis 
technique is provided by the Sloan Digital Sky Survey (SDSS, e.g., York et
al. 2000, Stoughton et al. 2002, Abazajian et al. 2003). 

In this paper we describe initial results of a program that exploits the 
unique combined capabilities of the ROSAT and Sloan Surveys to eventually
obtain a sample 
of $\sim10^4$ X-ray sources, which are not only optically identified, but which 
also have high quality and uniform data in the X-ray from RASS and in the 
optical (photometry and spectroscopy) from SDSS. 
Our emphasis in this initial 
paper is on high-confidence RASS/SDSS optical identifications 
of quasars and other AGN. 
In section 2 we provide a 
brief description of relevant aspects of the SDSS data, and the fortuitous
match in sensitivity to the RASS. In section 3, we 
detail the technique used to select and confirm candidate counterparts from 
the SDSS data. In section 4, we present results from this 
RASS/SDSS program for 1400~deg$^2$ of sky, providing an initial catalog 
(optical photometry and 
spectroscopy, as well X-ray properties) of more than 1200 likely AGN
identifications for RASS X-ray sources, including 45 BL~Lac
candidates. In section 5, we discuss a few additional AGN
subcategories and objects of special interest, including NLS1s. 
Section 6 provides an overview of the
ensemble properties of the initial sample, especially as they pertain to
the reliability of the identifications. Section 7 provides
an illustrative example of optical/X-ray correlation studies that may be
enabled by the sample. A very brief summary 
is provided in Section 8. 

\section{The Excellent Match Between SDSS Optical and RASS X-ray Surveys}

The Sloan Digital Sky Survey (SDSS) 
is a multi-institutional 
project to create an optical digital imaging and spectroscopic data bank of 
a large portion of the celestial sphere, mainly in a region approaching 
$\sim10^4$~deg$^2$  centered on the north Galactic polar cap. A recent 
overview of the project has been given by York et al. (2000), and a 
comprehensive technical description of the large subset of the software 
pipelines and data released earlier to the public is available in 
Stoughton et al. (2002). The optical data are obtained by a special 
purpose 2.5m telescope,
located at Apache Point Observatory, New Mexico, built and optimized solely for this 
project. It is equipped with a unique mosaic camera 
(Gunn et al. 1998) which operates in TDI (Time Delay and Integrate) mode
that can image $\sim10^2$~deg$^2$ in 5 colors in a single night, as well 
as a multifiber spectrograph which obtains the spectra of 640 objects 
within a 7~deg$^2$ field simultaneously.  The resulting photometric atlas is
in 5 specially-chosen colors $u,g,r,i,z$ (Fukugita et al. 1996,
Hogg et al. 2001, Smith et al. 2002)
from $\sim$3600 to 10,000~\AA, covering the entire
survey area to a limiting magnitude of $r$$\sim$22--23, 
on $0.4''$ pixels. This database is automatically analyzed to
catalog the photometric 
and astrometric properties 
of several hundred million stars and galaxies, as well as a million quasars.
The imaging database is used to autonomously and homogeneously select 
objects for the SDSS spectroscopic survey, which will 
ultimately include moderate resolution ($\lambda/\Delta\lambda\sim1800$) 
spectrophotometry covering a broad (3800-9200\AA) wavelength regime
for $10^6$ galaxies, $10^5$ quasars, and $10^5$ unusual stars. 

The RASS and SDSS are extremely well-matched to each
other via a variety of fortunate coincidences. For example, the SDSS survey
area is also the region of greatest RASS X-ray sensitivity, because it
lies primarily at high ecliptic latitude (and thus long RASS
exposure times) and also high Galactic latitude (and thus regions of low
interstellar photoelectric opacity, an important effect in the
soft ROSAT energy band).  

Perhaps most important, the limiting X-ray flux of
RASS is such that, given the known range of $f_x/f_{opt}$ for most common
classes of sources, both Galactic and extragalactic, even the optically
faintest counterparts of typical RASS sources are accessible to both the 
SDSS photometric and spectroscopic surveys.  For example, although there is
considerable dispersion, the most X-ray luminous (relative to optical)
normal stars, normal galaxies, quasars, and BL Lacs are known to have
$log~(f_x/f_{opt})$ of about  -1, -0.25, 1.0, and 1.5 respectively ({\it
e.g.}, Stocke et al. 1991). At the limiting RASS X-ray flux quoted above,
this implies that even unusually faint optical counterparts in each of these 
classes will have magnitudes brighter than  $m\sim$ 15, 17, 20, and 21,
respectively. Thus the
SDSS imaging survey will obtain highly accurate colors and magnitudes for
the vast majority of RASS counterparts, and even the spectroscopic portion
of SDSS, where the $\sim$45~minute exposure times are set primarily to 
provide acceptable signal
for the galaxies in the large scale structure survey, will yield excellent
quality spectra for the large majority of counterparts, permitting confident
identifications and in-depth subsequent analyses.

Additionally, the full RASS/SDSS area is also covered by the NVSS and/or
FIRST 20~cm radio surveys (Condon et al. 1998; Becker, White \& Helfand 1995). 
Thus radio information is also available for virtually {\it every}
interesting RASS/SDSS object. 

The final part of the good match between RASS and SDSS is a consequence of the
modest surface density of RASS sources on the sky---commonly, a few sources per 
square degree.  A typical SDSS spectroscopic field of about 7~deg$^2$ thus has 
$\sim10^1$ RASS sources, and therefore $\sim10^4$ RASS sources in the ultimate 
SDSS sky coverage can be targeted for at least one spectroscopic observation, 
while only diverting 1\% of the fibers from other SDSS scientific programs.

\section{Autonomous Selection of Candidate Optical Counterparts for SDSS Spectra}

As noted above, a subset of objects of special interest in the SDSS imaging 
are slated for follow-up SDSS spectroscopy; objects are selected from
the imaging database via various algorithms that comprise
the SDSS ``target selection" pipeline software (Stoughton et al. 2002).
As part of that target selection pipeline, we have implemented
algorithms aimed to specifically select candidate ROSAT counterparts for
automated SDSS spectra. Optical objects in the SDSS photometric catalogs 
are automatically cross-correlated with objects in the RASS X-ray catalogs
(Bright and Faint Source).
Those SDSS optical objects within $1'$ of the X-ray source positions are 
initially flagged (with a database flag `rosatMatch' having
value $>$0) as potential 
positional matches in the imaging catalogs and considered further, though 
not all will receive follow-on SDSS spectra as we detail below.

As our eventual goal is the optical identification of $\sim10^4$ RASS X-ray 
sources (a sample approximately an order of magnitude larger than previous
efforts), we emphasize SDSS spectroscopic 
follow-up of that subset of the cataloged RASS sources which have
the highest X-ray detection 
likelihoods (a maximum likelihood measure of the RASS X-ray source 
significance; see Voges et al. 1999). In the RASS/SDSS target
selection algorithm, the minimum X-ray detection likelihood parameter 
is currently set to $\ge10$; this cut, imposed on the RASS catalog,
yields the desired highest-significance $\sim10^4$ X-ray sources in the
ultimate joint RASS/SDSS sky coverage.

While it would be optimal to take optical spectra of all objects within 
each such RASS error circle, this is not feasible for several reasons in 
practice. First, the minimum allowed spacing between SDSS spectroscopic 
fibers is 55$''$, and most error circles will be sampled spectroscopically
on only a 
single occasion during the course of the entire SDSS survey. 
Second, the SDSS imaging data extend to beyond $m\sim22$, and therefore the faintest 
optical images detected are significantly fainter than the SDSS 
spectroscopic limits. A compromise is made to 
generally aim for taking an SDSS spectrum of a single optical object 
in each RASS error circle (with X-ray maximum-likelihood value $\ge10$)
covered by SDSS. The optical object chosen for SDSS spectroscopy in each
RASS circle must also be brighter than 
$m<20.5$ (more specifically, $g$, $r$, or $i$ $<20.5$); this is the
magnitude limit for
good quality SDSS spectra of emission line objects in the typical 
45~minute spectroscopic exposure
times. There is also a bright limit of $m>15$, needed 
to avoid spectroscopic cross-talk between nearby multi-object fibers.

An additional complication is that, for technical
reasons, photometric measures for all SDSS optical 
objects within each RASS circle may
not generally be {\it  intercompared} to one another at the stage when 
algorithmically selecting targets for SDSS optical fiber 
spectroscopy.\footnote{For example, in an extreme case it could even be that
early SDSS optical imaging covers only some portion of a given
X-ray error circle, while optical imaging coverage of the remainder of the
X-ray error circle will not be completed until very late in the survey.}
Thus, rather than using the
SDSS multicolor imaging data to select the ``best" candidate optical
counterpart in each RASS circle for a spectrum, we instead
(at the target selection stage) must
independently assess the quality of each optical object in 
the RASS circle on an absolute scale, without knowledge of 
(or direct comparison to) any other positionally consistent optical objects.

Our approach is thus to assign every SDSS optical object (imaged 
at the time of ``target selection") within the
RASS circle to a broad priority bin grade (A,B,C,D)
for optical spectroscopy. These priority 
bins are based on 
typical ratios of X-ray to  optical flux for various classes of known X-ray 
emitters, at a typical limiting RASS X-ray flux of a few times $10^{-13}$ 
erg/sec/cm$^{2}$, and are used to gauge 
identification likelihoods and hence assign spectroscopic priorities for SDSS 
candidate optical counterparts. SDSS optical magnitudes, colors and 
morphology parameters, as well as FIRST radio data, serve as proxies 
at this stage for the likely  object class. 

Among the various possible priority bins for ROSAT counterpart
identifications, we have chosen objects
having a triple positional coincidence between a RASS X-ray source,
an SDSS optical object, and a FIRST radio source, for highest priority for
SDSS optical spectra; the SDSS object and FIRST source must match to within 
$2''$ (which we note may miss some radio sources with complex morphology). 
As mentioned above, virtually
100\% of the SDSS area will also be covered by the FIRST 20~cm radio survey, 
although the relevant FIRST catalogs may not always have been
available when the SDSS target selection software was actually run. 
The cross-correlation with FIRST catalogs (also done automatically)
is mainly intended to ensure sensitivity 
to BL~Lacs and other X-ray/radio-emitting quasars, but this algorithm also 
finds, for example, 
radio galaxies (isolated and in clusters), etc. These highest-priority 
triple-waveband RASS/SDSS/FIRST coincidences are 
flagged as `ROSAT\_A' objects by 
the target selection pipeline software, and indicated as
such in the SDSS database.

Objects in RASS error circles with unusual SDSS colors 
indicative of AGN and quasars (but lacking FIRST detections)
are given next priority for SDSS optical spectra. 
Simple UV-excess selection $u-g<0.6$ built into the 
RASS/SDSS target selection module,
as well as far more sophisticated approaches (Richards et al. 2002;
Newberg \& Yanny 1997) looking for outliers in the full
SDSS 4-dimensional color space, are consulted in 
assigning objects to this potential AGN class. All such second-highest 
priority potential spectroscopic targets are flagged as `ROSAT\_B' by the 
target selection algorithm module. In fact, more than 90\% of the 
confirmed RASS/SDSS quasars/AGN are selectable by the simple UV-excess criterion
(see Figure~1);
this is not surprising, as X-ray selection at the RASS flux limits has
long been known to favor low-luminosity quasars and related AGN at low-redshift
(e.g., Margon, Chanan, \& Downes 1982). These color-outlier selections 
(especially because of the sensitivity to UV-excess) also 
sometimes yield identifications which, upon examination of
the SDSS spectra, turn out to be X-ray emitting cataclysmic 
variables, white dwarfs, etc. (The latter classes are
the topics of other SDSS papers).

The third broad priority bin includes a varied set of somewhat less 
likely or less interesting, but still plausible, identifications including: 
bright stars and galaxies (currently $r<16.5$ and $r<17.5$, respectively;
but also recall the general spectroscopic {\it bright}
limit of $m>15$), unusually blue galaxies ($g-r<0.2$), and moderately 
blue stars ($u-g<1.1$).
These medium priority spectroscopic targets are flagged as
`ROSAT\_C' by the target selection pipeline module.

The fourth broad priority bin includes any object bright
enough (but not too bright) for optical spectroscopy, 
that falls within the RASS circle. Objects in this class are flagged
as `ROSAT\_D' by the target selection pipeline, and most are rather unlikely to 
be the proper identifications; on the other hand, it is important to 
allow for this group as well, lest our biases based on past studies preclude 
discovery space for new classes of X-ray emitters.

Among the confirmed quasars/AGN cataloged in this paper,
selected for possible SDSS spectra by virtue of their interest as 
potential RASS identifications, the great majority were (as expected) 
assigned to either--or sometimes both--`ROSAT\_A' and `ROSAT\_B' categories. 
When considering just the highest priority bin assigned: 16\% were assigned 
`ROSAT\_A',  76\% were assigned `ROSAT\_B', 3\% `ROSAT\_C',
and 5\% `ROSAT\_D'. 

One subtlety of RASS/SDSS target selection is that 
because of the restriction noted above that optical
objects within an X-ray error circle may {\it not} have their
photometry inter-compared beforehand to pick the ``best" candidate, a
situation may still arise in which, for example, both
a `ROSAT\_A' and `ROSAT\_D' object, well separated in the
same X-ray error circle, could otherwise still both be assigned
an SDSS spectroscopic fiber. In order to avoid wasting the second 
fiber in such a case, a trick is now imposed that 
effectively restricts spectroscopy in most cases to only a single SDSS 
candidate
in each X-ray error circle: currently, only optical candidate counterparts 
within radius $27.5''=55''/2$ offset from the X-ray source are ultimately
allowed for selection for spectroscopy by the target selection
software module devoted specifically to RASS/SDSS.
The minimum $55''$ mechanical spacing between spectroscopic fibers
then ensures that only a single SDSS optical object within the
X-ray error circle will actually receive a RASS/SDSS spectroscopic fiber:
one in the highest priority bin offset $< 27.5''$ from the
X-ray source. (If there are multiple SDSS objects
of the same priority---e.g., 
two SDSS objects within $27.5''$ of a RASS source, both having priority 
class `ROSAT\_A'---then one is chosen at random for SDSS spectroscopy).

For aid in database book-keeping, 
we have added an additional flag `ROSAT\_E',
that is now also assigned at the target selection stage 
to indicate an SDSS optical object of potential interest that is 
unsuitable for follow-up SDSS spectroscopy for any reason. 
For example,
this flag might call attention to an SDSS object of interest 
outside the $27.5''$ radius 
offset discussed in the preceding 
paragraph (but within $1'$ of the RASS source position), or one too 
bright or too faint for SDSS spectroscopy, 
or one matching a RASS source with X-ray detection likelihood 
below 10, etc. (Note also that an SDSS object may receive more than one
ROSAT target selection flag, e.g., one with `ROSAT\_A' and `ROSAT\_E'
flags set might indicate a triple radio/optical/X-ray match, but one
not suitable for SDSS spectra for some reason, etc.).
We added the `ROSAT\_E flag during the early phases of SDSS commissioning,
but after early target selection was already underway; however, this
late addition complicates the book-keeping on only about 10\% of the objects 
discussed in this paper (and of course that fraction will decrease, as the
sample expands).

The actual SDSS spectroscopic targeting situation is even somewhat
more complex than described here (see Stoughton et al. 2002
for additional details). First, the target selection 
algorithms for each of the various other SDSS spectroscopic follow-up
categories---quasars, galaxies, stars, serendipity objects, as
well as ROSAT candidates---were under refinement in the 
early ``commissioning" phases of SDSS.
Secondly, objects that are part of the SDSS optical samples to obtain 
(ultimately) $10^{6}$ spectroscopically-observed galaxies 
and $10^{5}$ confirmed quasars are actually assigned their
spectroscopic fibers first via a sophisticated ``tiling
algorithm" (Blanton et al. 2003), before {\it any} consideration
is given to potential RASS targets. For example, an SDSS object
selected as a QSO candidate by the {\it quasar} target selection algorithm
will very likely receive a spectroscopic fiber (and e.g., even when the
ROSAT algorithm flags it as `ROSAT\_E').
Indeed, the following SDSS optical
objects are nearly guaranteed to receive an SDSS spectroscopic fiber
independent of X-ray emission: all galaxies to $r<17.77$ (Petrosian
magnitude), almost all SDSS PSF objects to $i<19.1$ with 
colors indicative of quasars, and almost all FIRST radio sources with 
optical PSF-morphology to $i<19.1$.
A common benefit is that the number of distinct SDSS 
fibers needed specifically for RASS sources is thereby 
further decreased due to the large fraction---about 80\%---of RASS 
identifications that ultimately prove to be AGN,
and which were already independently 
targeted for spectroscopy 
by the SDSS {\it quasar} target selection algorithms
(before the algorithms
specifically aimed at RASS candidates ever come into play).
Moreover, the {\it quasar} target selection algorithms are 
not confined to pick only objects within the $27.5''$ offset
restriction imposed on RASS/SDSS targets, so additionally can sample
counterparts throughout the full $1'$ X-ray error circles.
Conversely though, a random galaxy targeted at high priority merely as it is 
part of the main SDSS galaxy redshift-survey sample, and falling by chance 
in a RASS error circle, may occasionally {\it exclude} another 
high-priority ROSAT target from any possibility of receiving a 
spectroscopic fiber. This is merely due to the $55''$ minimum fiber spacing. 

The net effect is that RASS/SDSS target selection, both in general 
and also specifically for the objects cataloged in this paper,
cannot guarantee ``completeness". However, in practice we expect
the spectroscopic sample will ultimately prove to be reasonably complete 
at least for most classes of
X-ray emitting quasars and AGN with $15<m<19$. Within $27.5''$ offset
from the RASS sources, such optical targets may be chosen for 
spectroscopy by both the quasar 
and the ROSAT target selection categories (and at
larger positional offsets, about 80\%
of X-ray emitting quasars/AGN like those discussed
herein may still be selected by the {\it quasar}
target selection algorithms.)

Despite the limitations on completeness, examination of the very 
high-quality SDSS spectroscopy (e.g., see Figure~2) confirms that these target 
selection algorithms that choose
odd-colored optical objects and/or radio sources within RASS error circles 
are highly efficient at locating X-ray emitting quasars/AGN. 
For example, objects selected by the `ROSAT\_A' and/or `ROSAT\_B' algorithms 
applied to the SDSS imaging data are confirmed to be X-ray emitting 
AGN conservatively more than about 75\% of the time, upon 
examination of the follow-up SDSS spectra;
in fact, approximately 85\% of the spectra in these two
algorithm categories are those of very likely 
quasars/AGN (and constitute the catalogs presented herein), but a few 
percent of these may be chance superpositions unrelated to 
the X-ray sources as we discuss in section 6, and a few percent 
reflect slightly more speculative spectral classifications/identifications (the
Sy~2 candidates discussed near the end of section 4.2, a handful of objects
among the BL~Lac candidates discussed in section 4.3, and a few dozen
of the objects among the large sample of NLS1s discussed in section 5.1).
The efficiency of spectroscopic identifications may also be somewhat 
larger 
than quoted above for the `ROSAT\_A' and `ROSAT\_B' categories, as these 
same algorithms also yield: 8\% other sorts of galaxies (4\% are galaxies 
with comparatively weak/narrow or no optical emission, plus 4\% radio 
galaxies 
having bland optical spectra); 3\% stars, most of which are either white dwarfs 
or cataclysmic variables; and, 4\% of the spectra that are either 
ambiguous or of too low S/N to definitively classify. Some fraction of these 
additional objects (e.g., nearly {\it all} of the CVs---which are 
already cataloged 
as likely X-ray sources in other SDSS papers by Szkody et al. 2002, 
2003) are also plausible additional RASS IDs, but are not included, nor 
counted as successes, in the discussions of the current AGN paper.

\section{Initial Catalog of 1200 X-ray Emitting AGN From RASS/SDSS}

In this section, we present initial results towards the ultimate goal of 
a sample of $\sim10^4$ well-characterized optical identifications of
RASS X-ray sources. The emphasis in this installment of the
RASS/SDSS catalog is on our initial set of more than 
1200 X-ray emitting quasars 
and other sorts of AGN; as discussed below in section 6, for these 
AGN the identifications are secure at least in a statistical sense. The
comprehensive range of Galactic and extragalactic X-ray identifications
will be discussed in upcoming
papers related to the SDSS Data Release 1 and beyond (e.g., Voges et al.
2003).
Other ROSAT results especially from the SDSS Early Data Release (EDR),
bearing on various specific X-ray emitting classes include:
cataclysmic variables (Szkody et al. 2002, 2003), 
clusters of galaxies (Sheldon et al. 2001), 
EDR narrow-line Seyfert 1s (Williams, Pogge, \& Mathur 2002), and EDR quasars 
(Schneider et al. 2002; Vignali, Brandt, \& Schneider 2003a).

\subsection{Quasars and Other AGN with (Predominantly) Broad Permitted Lines}

In Tables 1 and 2 we present a catalog of 964
spectroscopically confirmed RASS/SDSS
quasars and closely-related AGN having (predominantly) broad permitted
emission lines; all have optical positions
within $1'$ of RASS X-ray sources. Sample Tables 1 and 2 (listing
just the first 5 entries) are included within
this paper itself; the full tables are available electronically 
from the Journal, or upon request to the authors 
(contact the lead author).
We include under this present category of quasars and other ``broad-line"
AGN all such RASS/SDSS objects whose
spectra show characteristic strong optical emission lines of AGN, with
broad permitted emission having velocity width in excess of 
1000~km/sec FWHM. For low redshift objects in the 
range out to about $z<0.4$
where MgII 2800 is not yet accessible, we rely on the $H\beta$ 
emission line to make this velocity-width determination;
for higher-redshift AGN, we merely consider the FWHM of
the broadest line observed within the SDSS spectroscopic coverage.
Note that, aside from luminosity and data quality
criteria, this category is defined similarly to that of the Schneider et al. 
(2002) EDR quasar catalog. However, we here avoid any {\it a priori}
luminosity cuts, and so these first tables will include not only most quasars, 
but also most Sy~1s, many Sy~1.5s, and even many of such unusual but 
interesting classes as NLS1s. 

Nearly all (about 98\%) of these ``broad-line" AGN were selected for 
spectroscopy by either the ROSAT or 
the quasar target selection algorithms (or both as discussed above) applied
to the SDSS imaging database.\footnote{As an additional check on
the autonomous target selection algorithms discussed above, we also have
considered {\it post facto} (and cataloged here as AGN where appropriate) all 
SDSS objects with an optical spectrum that fall within 1$'$ of
a RASS source, and independent of whether or not actually selected by 
the ROSAT target selection algorithms described in the preceding section.
For this additional check we externally cross-correlated with the RASS catalogs 
the positions 
of $\sim 200,000$ optical objects having spectral information incorporated into the 
SDSS database as of early 2002, and examined all 3141
resulting SDSS spectra.
This additional check confirms that the autonomous RASS/SDSS target selection 
pipeline
algorithms are working as anticipated, and are not missing any substantial 
set of X-ray emitting AGN.}
This catalog reflects objects having follow-up SDSS spectroscopy 
parameters stored within a database accessible to the SDSS collaboration 
(for internal reference, we refer to this as the ``Chunk 8 database") 
as of early 2002; that database included information from
269 (distinctly numbered) spectroscopic plates, of which 251 are in common
with spectroscopic plates included
in the most recent public data release (``Data Release 1";  Abazajian 
et al. 2003).
All objects have confirming, high-quality SDSS 
optical spectra; see Figure~2 for representative examples.
About 20\% of these objects were both already-cataloged as quasars/AGN 
and also had redshifts reliably established prior to
SDSS, and the bulk are newly-identified as X-ray emitting AGN
from combined RASS/SDSS data. Other objects
are previously cataloged in Bade et al. (1998a) and related lists,
but lacked secure identifications in their low-resolution grism/prism 
spectral data.
(See also Margon et al. 2000, Voges et al. 2001, 
Schneider et al. 2002, Williams et al. 2002, and Vignali et al. 2003a for 
discussions of various subsets of the several hundred AGN identifications available 
earlier from the EDR.) In this paper, Table~1 catalogs mainly empirical
characteristics of the 964 X-ray-emitting broad-line quasar/AGN counterparts 
obtained from SDSS, while Table 2 provides other derived information discussed 
herein. In both tables, objects are ordered according to the
J2000 RA of the RASS X-ray source.

In Table~1, emphasizing observed parameters, 
the {\it 1st column} provides the RASS X-ray source
position (J2000) using RA/Dec nomenclature. The {\it 2nd column}
similarly provides the optical position/nomenclature (J2000) of 
the suggested quasar/AGN
counterpart as measured from SDSS data; the
optical astrometry is accurate to better than $0.1''$ (Pier et al. 2003).
The {\it 3rd through 7th columns} provide optical photometry
in the 5 SDSS passbands (e.g., Fukugita et al. 1996) in the
{\it asinh} AB system (Lupton, Gunn, \& Szalay 1999); as the final
SDSS photometric calibrations were not completed at the time 
this project began, the
PSF magnitudes in Table~1 are denoted with an asterisk,
and the estimated calibration uncertainties
are of order a few percent. 
The {\it 8th column} provides the value of the SDSS imaging parameter
`objc\_type', a commonly used and reliable 
SDSS measure of optical morphology (see Stoughton et al. 2002 for 
more details);
objc\_type=6 indicates stellar/unresolved optical morphology, while
objc\_type=3 indicates an extended/resolved (i.e., galaxy) morphology.
The {\it 9th column} provides the redshift as
measured from the SDSS spectra; in most cases the quoted redshift
is that provided by the SDSS spectroscopic pipeline,
but we have confirmed all redshifts by an independent manual/by-eye check,
and agreement in most cases (except for a handful that we have
corrected in the tables based on our manual measures) was to within 
0.01 in~$z$. The remaining columns of Table~1 emphasize the X-ray data on the 
broad-line quasars/AGN, and are derived
directly from the RASS catalogs (e.g., see Voges et al. 1999, 2000). 
The {\it 10th column} provides the RASS X-ray source
count rate (counts/sec) in the 0.1-2.4~keV broadband, corrected for 
vignetting. The {\it 11th column} gives the RASS exposure time in seconds.
The {\it 12th and 13th columns} provide
X-ray hardness ratios as measured from several X-ray bands
(see Voges et al. 1999).
The {\it 14th column} is the X-ray source detection likelihood (a maximum
likelihood measure of source significance). The {\it 15th column}
gives the estimated X-ray flux in the 0.1-2.4~keV band, 
as observed, i.e., without 
corrections for absorption in the Galaxy; we used
the PIMMS (Portable, Interactive, Multi-Mission Simulator) 
software to convert RASS count rates 
into X-ray fluxes, assuming
a power-law X-ray spectrum with energy index $\alpha_x=1.5$ typical of
(low redshift) quasars in the RASS passband (e.g., see Schartel et al.
1996).

In Table 2, we present further catalog information on the 964
broad-line RASS/SDSS AGN, but now also emphasizing derived quantities
used herein.
For more convenient cross reference we repeat some relevant information
included within Table 1. The {\it 1st column} and {\it 2nd column} repeat,
respectively, the RASS X-ray source and suggested optical counterpart
name/position.
The {\it 3rd column} provides the extinction-corrected
$g$-band PSF magnitude, 
corrected according to the reddening
maps of Schlegel, Finkbeiner, \& Davis (1998). 
The {\it 4th column}
repeats the redshift from SDSS spectroscopy.
The {\it 5th column} is the X-ray flux (units of
$10^{-13}$~erg/s/cm$^2$) in the 0.1-2.4~keV band, now
corrected for absorption within the Galaxy,
and with absorbing columns estimated from the $N_H$ column density
measures of the Stark et al. (1992) 21cm maps.
The {\it 6th, 7th, and 8th 
columns} give the logarithms of, respectively, inferred broadband
(0.1-2.4~kev) X-ray luminosity (units of erg/s), monochromatic
UV/optical luminosity (units of erg/s/Hz) at a frequency 
corresponding to rest frame 2500~\AA\ , and  monochromatic
soft X-ray luminosity at 2~kev; we adopt 
values of $H_0$=70~km/sec/Mpc, $\Omega_M=0.3$, and
$\Omega_{\Lambda}=0.7$ for deriving the luminosities in Table~2.
In converting from corrected broadband (X-ray 0.1--2.4~kev, and optical
$g$-band) flux to luminosities, we again
assume an X-ray power-law spectrum with energy index $\alpha_x=1.5$,
and  an optical power-law with energy index $\alpha_o=0.5$.
The {\it 9th column} lists $\alpha_{ox}$, the 
slope of a hypothetical power law (in energy) from the UV/optical to X-ray 
(i.e., connecting 2500~\AA\ and 2~kev). 
The {\it 10th column} provides brief
comments, e.g., noting selected objects that are radio sources,
including one alternate name for 
objects whose spectral classification and accurate redshift
were both available (in catalogs/papers incorporated into NED) prior to SDSS,
noting the $\sim$1\% of the cases where two of our cataloged AGN fall 
within the same RASS error circle 
(latter denoted as 'ambigID' in the comments), etc. 
Again, only the initial 5
entries are included in this sample Table 2 within this paper itself; the full
table is available electronically from the Journal and the lead author, 
as indicated above.

\subsection{X-ray Emitting AGN Having Narrower Permitted Emission}

In addition to the predominantly ``broad-line" (FWHM$>$1000~km/sec)
AGN discussed above in 
section 4.1,  we also examined and approximately classified the spectra 
of all other extragalactic emission line objects in RASS error circles
observed spectroscopically by SDSS.
Primarily, this was done to ensure fuller sampling of the 
specific possible X-ray emitting AGN subclasses like NLS1s,
Sy~1.5s, 1.8s, and 1.9s that have
optically-observable broad-line regions; i.e., objects very closely 
related to classic quasars and Seyfert 1s, but which may be observed to 
have narrower permitted line components as well. In addition, we
have also extended this subset to include Seyfert 2
candidates, i.e., AGN whose optical emission is nearly entirely dominated by
a narrow-line region.
The additional AGN cataloged in this section are
exclusively at low redshift, so again it is
initially the H$\beta$ line width/profile that we consult. That is, the objects
discussed in this section all have FWHM(H$\beta$)$<$1000~km/sec based
on a {\it naive} (and in many cases here, incorrect) line measure that 
assumes a single component to the line profile. 

For classification
as Sy~1.5 to Sy~1.8, we require the full width near
the continuum level (hereafter, FWZI) of the H$\beta$ emission line 
to exceed 2500~km/sec. 
Of course measurements near the continuum are problematic near H$\beta$, 
especially when Fe is strong, but except for the unusual case of some NLS1s, 
we have also verified that the FWZI
of the H$\beta$ line exceeds that of the 
O[III] 5007 emission in the same object by at least $\sim$1000~km/sec; hence 
each such object has a ``broad-line" component substantially wider 
than typical of its narrow line region. 
The taxonomical classifications provided in this section are mainly
intended as an aid in clarifying why these objects, although similar, did not satisfy 
the specific criteria discussed in section 4.1 for (predominantly)
``broad-line" quasars and AGN. To make the specific taxonomical subdivisions 
simple but more concrete, we label as Sy~1.5 those objects having 
{\it both} broad and narrow H$\beta$ components and with relatively small 
[OIII]~$\lambda$5007 to H$\beta$ flux ratios of $R<3$, and label
those with $R>3$ as Sy~1.8s; see Whittle (1992) for a similar
sub-classification scheme.\footnote{We do not attempt here to 
sub-classify cases like Sy 1.2s, intermediate between Sy 1 and Sy 1.5.}
(The NLS1s are discussed
in more detail in section 5.1). 

We also select a number of low redshift 
objects in RASS error circles, whose SDSS optical spectra
show little or no evidence for broad H$\beta$, but which do appear to have 
a markedly broad H$\alpha$ emission component. We refer to these
(somewhat loosely) as Sy~1.9s; however, in low S/N cases, 
weak broad H$\beta$ may not be well-measured or even recognized,
so some of these are likely to really be Sy~1.5--1.8s. To help limit 
confusion in these ``Sy~1.9" cases between truly broad H$\alpha$ and 
those in which narrow H$\alpha$ is merely blended with
[NII] emission, as well as complications of contaminating late-type
stellar continuum/absorption features in cases with substantial host-galaxy 
stellar light, we very conservatively limit our Sy~1.9 candidate list
here to include just those cases with very broad H$\alpha$,
requiring FWZI(H$\alpha$)$>$6000~km/sec.

Finally, we also extend our catalog a little to include candidate 
X-ray emitting Sy~2s and related AGN. Possible identifications of Sy 2s, 
starbursts, and 
other narrow emission line galaxies as X-ray emitting subclasses have been 
discussed in multiple papers most recently emphasizing deeper, higher 
spatial-resolution, and/or harder energy X-ray imaging surveys 
(e.g., Boyle et al. 1995, McHardy et al. 1998). However, in some cases 
subsequent improved follow-on optical spectroscopy has led to 
ultimate reclassifications (e.g., Halpern, Turner, \& George 1999) into one of the categories 
already considered above, e.g., NLS1, Sy 1.8, Sy 1.9, etc. Nonetheless, an 
extension to include X-ray emitting Sy 2 candidates may be of interest for
several reasons including: Type 2 AGN are sometimes invoked as significant
contributors to the cosmic X-ray background, and nearby bright examples
may be individually interesting if {\it bona fide} Sy 2s; and, even if
the candidates discussed herein are not confirmed 
as X-ray emitting Sy 2s upon closer scrutiny, this group might still 
serve to encompass some unusually subtle cases of X-ray emitting  
Sy~1.8-1.9s, NLS1s, etc.
We select candidate Sy 2s from among the remaining narrow emission line
galaxies, using AGN emission line diagnostics that are founded on 
the oft-used Baldwin, Petersen, \& Terlevich (1981; hereafter, BPT) diagrams and 
criteria. We employ the specific criteria of Kewley et al. (2001) 
based on the relative line strengths of [OIII]$\lambda5007/H\beta$, 
[NII]$\lambda6583/H\alpha$, and S[II]$\lambda6717,6731/H\alpha$; the
Kewley et al. criteria quantify regions in the $O[III]/H\beta$ vs.
$[NII]/H\alpha$ and $O[III]/H\beta$ vs. $[SII]/H\alpha$ BPT diagrams populated
by AGN versus starbursts.\footnote{Except that we do not consider the 
relatively weak $OI$ line in our Sy 2 classifications.
See also Zakamska et al. (2003) and Kauffmann et al. (2003)
who apply similar classification approaches to much 
larger {\it optical}
samples of possible SDSS Type 2 AGN.}

We catalog in Tables 3 and 4 basic empirical and derived information for 
the additional 216 X-ray emitting AGN discussed in this section
(194 with observed broad-line 
regions, plus 22 Sy 2 candidates). Tables 3 and 4 for
these additional AGN are analogous to Tables 1 and 2, respectively,
presented above in section 4.1; and again, only sample tables
are included within this paper, with the full tables available electronically.
In Table 4, the comment column may include the subclass
type, and we denote some especially uncertain classifications with a 
question mark (e.g., `Sy~2?').
Figure 3 shows selected 
SDSS spectra from objects in this second diverse set of X-ray emitting AGN
(and see section 5.1 for further details on NLS1s).

\subsection{BL Lac Candidates}

We have also examined our RASS/SDSS sample for BL~Lac candidates, which
(by the subclass definition) would, of course, not be found in the above samples
of AGN with prominent emission lines. Because of their rarity (only a few \% of 
the objects in current quasar catalogs), and unusual properties (nonthermal 
continua, with strong X-ray and radio emission and nearly featureless optical 
spectra, high polarization, marked variability, and possible ``negative"
evolution--e.g., Urry \& Padovani 1995, Morris et al. 1991), large and 
well-defined 
samples of BL Lacs are eagerly-sought, but until recently have proven 
extraordinarily difficult to assemble. Their rarity demands large areal sky 
coverage, and yet their unusual spectral energy distributions and lack of 
strong spectral features often render traditional quasar/AGN optical-selection
incomplete and/or inefficient for BL Lacs. Despite numerous attempts, 
until recently there were only a handful of homogeneous BL~Lac samples of 
even modest-size from which to study their remarkable characteristics.

Many recent efforts (e.g., Stocke et al. 1991, Perlman et al. 1996, 
Laurent-Muehleisen et al. 1997, Bade et al. 1998b)
to obtain well-defined BL Lac samples utilize a 
combination of multiwavelength information from the X-ray, optical, and
radio bands to obtain high selection efficiency. Stocke et al. (1991), 
Perlman et al. (1996), and others noted that BL Lacs have distinctive
multiwavelength flux-ratios; e.g., BL Lacs have unusually large $f_x/f_{opt}$ 
ratios compared to other AGN, and nearly all those cataloged that may 
confidently be called BL~Lacs are radio sources as well. We employ
these same general approaches and criteria within our `ROSAT\_A'
target selection algorithm (as discussed in section 3) to obtain an 
initial list of RASS/SDSS BL~Lac candidates. (We again also consider 
{\it post-facto} all
optical spectra of SDSS objects taken in the relevant RASS circles, to ensure 
our algorithms are not missing a significant subset.)

We find/recover 38 objects we consider as {\it probable} X-ray emitting RASS/SDSS 
BL~Lacs. For all these higher confidence cases:
(1) the SDSS optical object is within $1'$ of a RASS source; (2)~the SDSS
object is also a positional match to a radio source (e.g., conservatively taken 
as $<2''$ in this initial study for matches to FIRST sources, or $<20''$ for matches 
to other radio catalogs); (3) our measures from the SDSS optical spectrum
reveal no strong emission ($EW<4$~\AA); and, (4)~there is no 
CaII H\&K break/depression evident in the SDSS optical spectrum, or any such break
present must be only very 
weak (quantitatively, the measured ratio of the galaxy's flux redward of 
any break to that blueward is required to be less than $<1.33$, e.g., 
see Stocke et al. 1991, Dressler \& Schectman 1987). The latter criterion reflects 
the circumstance that X-ray selected BL~Lac sometimes have weak breaks, but not
entirely featureless optical spectra (due to sometimes more pronounced starlight 
contamination from the host galaxies); of course, many of the RASS/SDSS BL~Lacs do 
indeed have approximately classically featureless optical spectra, as we show below.
We also find/recover as {\it possible} BL~Lac candidates 7 additional objects that are 
within RASS error circles and which either: satisfy the first three criteria (X-ray/radio
sources with no strong optical emission), but which just barely miss 
on the fourth criterion (2 objects that have CaII break flux ratios of between 1.33
and 1.40); or, have too low S/N in their SDSS optical spectra to claim the BL~Lac 
spectral nature of criteria (3) and (4) with much confidence (3 objects);
or, which show approximately featureless SDSS spectra, i.e., that satisfy 
criteria (1), (3), and (4), but where a close match has not been made to a 
radio source (2 objects). 

In the 1400 deg$^2$ of sky considered here then, a total 
of 45 candidate optical counterparts satisfy these criteria as probable or possible 
BL~Lacs, and basic information is provided for them in Tables 5
and 6 (which are analogous to Tables 1 and 2 for emission line AGN).
Again, only the initial 5
entries are included in these sample tables; the full
tables are available electronically from the Journal and the lead author.
In the comment column of Table 6, we use `zunc' to denote selected cases 
where the redshift from SDSS is highly uncertain, and the redshift is not 
available from (or is different in) the literature.
(Of course, the quoted redshifts may also be lower limits on
the actual BL~Lac redshift, if the spectral absorption arises in foreground
material rather than in the BL~Lac host galaxy.)
If a redshift is not obtained from the SDSS 
spectrum, nor already available in the literature, we adopt
$z=0.3$ (near the median of others in the sample) for estimating
$\alpha_{ox}$ in Table 6; as this is essentially a ratio of luminosities, aside
from precise K-corrections, the values of $\alpha_{ox}$ should be
approximately correct in most cases. In Table 6, we denote
the possible but less certain BL~Lac candidates as `BL?' in the comment 
column. Example SDSS spectra of selected BL~Lac candidates
are shown in Figure 4.

About half of these objects have been previously-reported in the 
literature as BL~Lacs or BL-Lac-like objects, and some other objects
are previously cataloged in Bade et al. (1998a) and related lists,
but lacked secure identifications in their low-resolution grism/prism 
spectral data.
Even among the previously-cataloged BL~Lacs, SDSS provides
the first high-quality published optical spectra for many objects, and
new spectroscopic redshift estimates (admittedly
uncertain in some cases) for 9 of the previously
cataloged BL~Lacs. Where redshifts were available prior to SDSS, 
there is also usually good agreement with our SDSS measures, and in total 
2/3 of the full sample of 45 have spectroscopic redshifts estimated 
from SDSS spectra, and/or other slit-spectra already published in the 
literature.

A few of the suggested BL~Lac identifications may eventually
prove to be misidentifications where a radio galaxy
lies within an X-ray emitting cluster or other group of galaxies 
(e.g., Rector, Stocke, \& Perlman 1999).
Others may prove to be stars upon higher S/N or
resolution spectroscopic follow-up; but it is worth recalling that all but 
a couple are both X-ray and radio sources, and so are more likely to be BL~Lacs
than highly unusual X-ray/radio stars. As an additional precaution
in the latter regard, we have also considered proper motion
information. Collinge et al. (2003) are independently optically selecting 
SDSS BL~Lac candidates using lack of optical emission and lack of significant 
proper motions (via comparison with the POSS) as their principal criteria. 
None of our 45 RASS/SDSS
BL~Lac candidates has significant proper motion ($>$20~mas/yr), e.g., suggesting 
little contamination by nearby hot white dwarfs, etc.

Thus, it seems likely that most of the BL~Lac identifications are 
correct (see also section 6 below). In the ultimate RASS/SDSS joint sky 
coverage, we may thus predict with some confidence that we will 
be able to quantitatively define a sample of several hundred
X-ray BL~Lac candidates.

\section{Discussion of Other Rare Classes and Individual Objects of Interest}

\subsection{Narrow-line Seyfert 1s (NLS1s)}

Among our RASS/SDSS AGN (including objects discussed in both sections 
4.1 and 4.2),
we identify from the SDSS spectra a total of 133 likely, and 
another 36 possible, X-ray emitting NLS1s. These objects 
are denoted by `NLS1' (likely) or `NLS1?' (possible)
in the comment columns of Tables 2 and 4.
Gallo et al. (2003) are investigating this 
interesting subclass in detail from the current RASS/SDSS samples, and 
Williams et al. (2002)  have recently discussed about 45 of these same objects
independently selected from the EDR. The reader is thus referred to those far more 
extensive investigations.
But briefly, these objects have unusually narrow permitted lines, though in other aspects 
are more similar to Sy~1s than Sy~2s; in the X-ray, these objects often
show strong soft X-ray excesses, and marked variability.
A variety of explanations have been
suggested for NLS1s, including unusually 
low-mass black holes, higher accretion rates relative to Eddington,
nearly pole-on orientation, unusually thick broad-line regions, etc. 
(e.g., see recent reviews by 
Boller 2000 and Pogge 2000).

The 133 X-ray emitting examples we label as reasonably confident cases
all have low
[OIII]~$\lambda$5007 to $H\beta$ flux ratios (we very conservatively restrict confident cases
to $R<1$), H$\beta$ 
FWHM less than 2000~km/sec, and strong optical
Fe emission---representative 
characteristics of the NLS1 class  (e.g., Pogge 2000). The 36 less 
certain examples have either 
[OIII]/$H\beta$ ratios of $1<R<3$, weaker Fe, or lower quality and therefore
more ambiguous SDSS spectra, etc. Among the 1/3 of the 
confident cases overlapping
with Williams et al. (2002), we find  good ($\sim$85\%) agreement between our 
spectral classifications and theirs. Representative SDSS spectra for NLS1s are shown
in Figure~5.

\subsection{BALQSOs}
Initial ROSAT studies (Green et al. 1995), and
numerous subsequent follow-on studies, have shown that BALQSOs
as a class are weak emitters in soft X-rays. This
deficiency at soft energies is usually interpreted
as due to absorption of low energy
X-rays in BAL material of surprisingly high column
densities, typically inferred equivalent to
$N_H \sim 10^{22-23}$~cm$^{-2}$ (e.g., Green et al. 2001
Gallagher et al. 2002).
There are several {\it possible} BALQSOs or
quasars with mini-BALs
among the RASS/SDSS X-ray identifications discussed here, but 
these are not yet confirmed definitively.
Shown in Figures~6a and 6b are the SDSS spectra
of two such possible cases.

The numbers sampled thus far are tantalizing but conservatively still somewhat 
too small for definitive conclusions about the paucity of BALQSOS among RASS/SDSS X-ray 
identifications. For example, SDSS spectral coverage yields good 
information on CIV BALs at redshifts above about $z>1.7$, and our RASS/SDSS 
sample thus far includes about 50 such objects. If high-ionization BALQSOs 
comprise of order 10\% of the quasar population, and if there were no bias 
against them in soft X-ray surveys, we would have expected to have found of 
order a half-dozen such BALQSOs within the current sample. Results for
low-ionization BALQSOs are similar: low-ionization cases 
are probably present among only a few percent of the quasar population, so 
that once again only about a dozen might have been anticipated (even if ``normal"
in soft X-rays) within our current sample of about 450 RASS/SDSS quasars with 
$0.5<z<2.2$ (for which Mg~II BALs might be easily identified from the SDSS 
spectra). Although the results for BALQSOs are thus yet not quite 
definitive, future expansions of the RASS/SDSS sample---ultimately encompassing
$\sim 5\times$ as many objects as considered here---should allow
rather stringent constraints on the incidence of BALQSOs in soft X-ray 
surveys.

\subsection{Other Miscellaneous Interesting Subclasses/Objects}

In addition to NLS1s with strong optical Fe emission discussed in section
5.1, there are another 71 objects with strong Fe emission 
(especially optical, but sometimes in UV), but 
with permitted line-widths in excess of the usual maximum of
2000~km/sec considered applicable to the NLS1 class. We denote
these in Table 2 with `Fe' in the comments column.
Example SDSS spectra are shown in Figures 6c and 6d.

In addition, there are also a half-dozen intermediate Seyferts 
which might be termed Sy~1.5--1.8, 
but where the broad weak H$\beta$ component appears markedly asymmetric
(and/or complicated with Fe emission); Figure 6e shows 
one such example. There are additional objects that have strong and broad MgII 
typical of classic quasars, but which if at slightly lower 
redshift might have been classified as about Sy~1.8 based on the H$\beta$
line strength/profile; Figure 6f shows an example, and one
in which there appears to be both broad strong H$\alpha$ and MgII, but 
only rather weak broad H$\beta$.

There are many additional cases with unusual optical line profiles, 
including objects with very broad ($\sim$20000~km/sec FWZI) lines (Figure
6g),  as well as other AGN with multiple-peaked line components (Figure 6h). Many
of the latter objects are included in the detailed study of Strateva et al. 
(2003).

\section{Discussion of Ensemble Properties and  Reliability of
Identifications}

In the 1400 deg$^2$ of sky considered here, more than 1200
X-ray emitting quasars or other AGN are
identified as likely RASS counterparts, each with uniform
optical photometry and high-quality optical
spectroscopy from SDSS. As discussed above, the SDSS optical spectra not only
permit accurate redshifts, but also quantitative classification 
(based on widths and 
line-ratios) into various AGN subclasses. In this 
section, we return 
to a discussion of the ensemble properties of our sample.

The optical magnitudes (Figure 7a) for the quasars/AGN extend over the entire range
accessible to routine SDSS spectroscopy of $15<m<21$, though
the median of $g=18.7$ is typical of past related optical
identification efforts. Interestingly, even at the brightest end,
there are dozens of X-ray emitting quasars/AGN brighter than
17th magnitude not cataloged prior to SDSS.
The redshift distribution for the quasars is also shown in Figure 7b, and
again the median redshift of $z=0.4$ is typical of previous identification
work at similar X-ray flux levels. However, our initial RASS/SDSS AGN catalog 
already includes 23 suggested quasar counterparts at $z>2$,
including one possible identification even for a
redshift $z=3.56$ quasar.\footnote{The $z=4.9$ quasar 
SDSSp J173744.87+582829.5 mentioned in Anderson et al. (2001),
that falls within (though rather far out in) a RASS error circle 
is {\it not} confirmed as an X-ray source counterpart in {\it Chandra} 
follow-up; the X-ray source has been positionally pinpointed in
{\it Chandra} images by Vignali et al. (2003b) to fall well away 
(about 45$''$) from the high-redshift quasar.}

The distribution of offsets (Figure 8) between the RASS X-ray positions and 
the SDSS optical positions is approximately as expected if the
1200 AGN are indeed the proper identifications, at least in a statistical
sense. For example, as shown in Figure 8a, 84\% of the
SDSS quasars/AGN fall within 30$''$ of the RASS X-ray positions, in 
approximate agreement with independent expectations about the RASS positional
uncertainty distribution as a function of X-ray detection
likelihood (Voges et al. 1999); the
latter may be derived from Tycho stars also detected in the RASS.  
Figure~8b shows a slightly different representation of these positional
offset data, where now the offsets have been
normalized relative to the RASS positional error; the latter accounts for
the dependence of the RASS positional uncertainty on the X-ray source 
significance, total X-ray counts detected, etc. Figure 8b also confirms that
the RASS positional error estimates are reasonable ones, e.g., with 
more than 97\% 
of the suggested identifications found within an offset smaller than 
3~times that predicted from the 
RASS X-ray positional error.
The small number with larger offsets
(about 3\% of the sample) is also 
consistent with the approximate expected number of random coincidences
between SDSS quasars and the RASS error circles sampled, as we describe
below. 

We also show in Figure 9
one last related measure of the distribution of positional 
offsets between RASS X-ray sources and various extragalactic 
SDSS optical objects (having SDSS spectra). 
As one simple indicator of the association of various object subclasses with
RASS X-ray sources, we display histograms of the distributions of
the squares, $r^2$, of the positional offsets between the SDSS optical positions
and the RASS X-ray source positions; i.e., we count the fraction of
objects falling within equal area annuli offset from the RASS X-ray source positions.
For a chance superposition of SDSS objects within RASS error 
circles,
these histograms would be approximately flat with $r^2$.
We conservatively limit consideration (in Figure 9) to objects that
fall just within the $r<27.5''$ offset discussed in section 3; all
the ``target selection" algorithms for selecting 
objects for SDSS fiber spectroscopy may work within this 
smaller region, while for greater offsets the `ROSAT' algorithms may not 
select targets for SDSS spectra (and therefore could produce 
an artificial truncation in the distributions for larger $r^2$).
Figure 9 shows separate $r^2$ histograms for 
quasars/AGN with predominant broad-lines discussed in section 4.1
(upper left),
quasars/AGN with narrower permitted lines discussed in section 4.2
(upper right), as well as the BL Lac candidates discussed in section 4.3
(lower left); in each of these cases, the
histograms for the AGN are very strongly-peaked at small $r^2$ values, as expected if these
AGN are statistically the proper (i.e., with low contamination) X-ray
source identifications.

For comparison, the lower right panel of Figure 9 shows the $r^2$ histograms 
for other normal galaxies (with weak or no emission) within 27.5$''$ of 
RASS sources.
(We have excluded error circles where we
have already identified the probable X-ray counterpart as among the 1225 AGN 
discussed in sections 4.1-4.3, or where there is a 
previously known X-ray emitting cluster). 
Although this figure suggests the possibility of a very weak
statistical correlation between some of these galaxies and RASS X-ray
sources, it especially confirms that contamination by random superpositions 
is very high for these normal galaxy cases---contamination 
an order of magnitude higher than for the AGN/quasars.
Although the ensemble of such more normal galaxies may well be suitable 
for follow-on ``X-ray image stacking" or ``survival analysis" statistical studies 
appropriate for samples with a large fraction of X-ray non-detections, they are 
not considered further in the current paper whose focus is on secure
X-ray source identifications of AGN.\footnote{For example, one might consider
looking for hidden AGN among some brighter subset of such normal galaxies 
that also have 
small X-ray/optical positional offsets. Indeed, limiting consideration to 
$g<18$ and offsets $<27.5''$ yields an $r^2$ distribution for 
16 bright normal galaxies that is only a little less 
peaked than 
that seen for the AGN in Figure 9. However, these 16 galaxies are sufficiently 
bright in the optical that only 2 have $log(f_x/f_{opt})>0.1$ and therefore 
significantly exceed the $f_x/f_{opt}$ range observed for other normal 
galaxies 
(e.g., Stocke et al. 1991). The exceptions are a well-studied Virgo galaxy 
(NGC\,4636) with an extended X-ray halo (e.g., Jones et al. 2002), 
and a radio galaxy RX J1725.3+5255 (e.g., Bade et al. 1998a) that may also 
be in a group or cluster.}

A related estimate of the expected number of random coincidences of 
quasars/AGN with RASS X-ray sources further quantifies
the statistical reliability of the cataloged quasar/AGN identifications.
The surface density 
of SDSS {\it optically}-selected quasars is of order 15/deg$^2$, 
while the combined area covered by all RASS error circles considered 
in the 1400~deg$^2$ area of this initial sample is
less than about 4~deg$^2$. Thus, as the above
positional arguments also suggest, only a small fraction 
($<5\%$) of the 1225 proposed quasar/AGN counterparts are likely 
to be spurious random chance positional coincidences, for which the RASS X-ray source 
and SDSS quasar/AGN are in fact unrelated. 

The distribution of the ratios of X-ray to optical flux for 1158 
quasar/AGN identifications with observed broad emission line regions
(we exclude the Sy 2 and BL Lac candidates, as both classes are
known to have distinct flux ratios vs. broad-line AGN)
is also as expected for typical X-ray emitting quasars/AGN.
It has been long recognized that roughly as much
energy is emitted in the X-ray as in the optical bands, as is reaffirmed
by the empirical distribution (Figure 10a) of $f_{x}/f_{opt}$ for 
the AGN considered here. In estimating the
flux ratio $f_{x}/f_{opt}$, we
adopt for $f_x$ the (corrected) 0.1-2.4~keV X-ray flux 
(from Tables 2 and 4), and estimate the optical broadband flux
in a 4000-9000 \AA\ bandpass using the $g$-band PSF magnitudes
(again from Tables 2 and 4) and assuming an optical power law
in energy with index $\alpha_o=0.5$.
The observed distribution (Figure 10a) 
of $f_{x}/f_{opt}$ for RASS/SDSS quasars/AGN 
is very similar to that found for the EMSS quasars/AGN (see Stocke et al.
1991). 

Similarly, Figure~10b shows the distribution of $f_{x}/f_{opt}$ 
ratios for the 45 RASS/SDSS BL~Lac candidates discussed in section 4.3. 
Although distinctly
different than the distribution for the quasars/AGN in
Figure~10a, the BL~Lac distribution we find here
is very similar to that found in earlier X-ray
selected BL~Lac samples (e.g., again see  Stocke et al.
1991). The agreement of these $f_{x}/f_{opt}$ distributions with 
expectations for both quasar/AGN and BL~Lac classes, again further 
confirms that the great majority of the suggested RASS/SDSS AGN
identifications are in fact likely correct.

\section{Discussion of An Example Multiwaveband Correlation}

The large survey sample size and uniformity and quality of observational
data present a variety of opportunities
for follow-on detailed studies of correlations between optical and X-ray wavebands
for AGN. Although this paper primarily presents the 
initial catalog information, we also include
here one illustrative
example of the sample's potential use in studies of correlations 
between X-ray and optical wavebands.

As depicted in Figure 11, the 1158 X-ray emitting AGN with 
broad-line regions discussed in sections 
4.1 and 4.2 (Sy 2 and BL Lac candidates
are again excluded for reasons discussed above) also 
appear to show 
the very long-recognized non-linear correlation 
between optical and X-ray wavebands found in many earlier studies/samples
(e.g., Avni \& Tananbaum 1985).
Although ours 
is, of course, an
X-ray selected sample, the relationship we find (Figure 11a)
between X-ray and optical
monochromatic luminosity, $l_{x} \propto l_{opt}^{0.89\pm0.01}$, is in excellent
agreement with (though with smaller formal
errors than) that found in some 
optically-selected samples also studied via ROSAT (e.g., see Green et al.
1996; Vignali et al. 2003a). This correlation
is virtually unchanged 
when excluding the 12\% of sample objects that are radio-detected.
However, the relation does appear to change 
when separately considering lower versus higher optical-luminosity 
objects; if we divide the sample in two at the median
value $log(l_{opt})=29.63$, we find $l_{x} \propto l_{opt}^{0.82\pm0.03}$
for the half of the objects above the median $l_{opt}$, and $l_{x} \propto l_{opt}^{0.98\pm0.03}$
for the half below the median (see also the discussion in the last
paragraph of this section).

Similar consistent results are obtained when 
alternately regressing 
$\alpha_{ox}$ 
against optical luminosity. For the
same total ensemble of 1158 quasars/AGN with observed broad-line regions,
this regression yields  $\alpha_{ox} \propto l_{opt}^{0.041\pm0.005}$, 
equivalent as expected to $l_{x} \propto l_{opt}^{0.89}$.
Figure 11b depicts this latter relationship in a slightly different
form. The solid line is the  best fit regression relation (fit to all
1158 points), while the error bars show the mean and the standard error
in the
mean value of $\alpha_{ox}$, as well as the mean and standard deviations in optical
luminosity, when considering averages taken in various optical luminosity
bins. [Note that the vertical error bars reflect the very
large number of data points (typically 70-220) averaged together to
estimate $<\alpha_{ox}>$ in each bin.]

Figure 11b once again suggests the possibility of more complex dependences than
assumed in our simple regression of $\alpha_{ox}$ versus log($l_{opt}$)
(or in the equivalent log($l_x$) versus log($l_{opt}$) relation discussed
above). Moreover, the 
general sense of little dependence of $\alpha_{ox}$ on $l_{opt}$
at lower $l_{opt}$'s, but
a stronger dependence at higher $l_{opt}$
has also been suggested by some other studies
(e.g., Yuan et al. 1998a).
We caution however, that a detailed study would
necessarily include proper account of
selection biases, dispersions in the quantities used in the regression,
multiple possible parameter dependencies (e.g., on redshift and radio
luminosity), and other aspects which we have neglected here.
For example, disparate dispersions
in optical compared with X-ray luminosities can yield apparent, but spurious, 
non-unity slopes in such simple correlation analyses between
$l_{opt}$ and $l_{x}$ (e.g., Yuan, Siebert,
\& Brinkmann 1998b). 
Nonetheless, this illustrative example suggests that
the large size, uniformity and quality of optical and X-ray data,
and range of physical parameters covered by the RASS/SDSS 
sample are well-suited to a variety of more detailed follow-on 
multiwaveband correlation studies.

\vspace{-3mm}
\section{Summary}

The SDSS optical and RASS X-ray surveys are well-matched
to each other via a variety of fortunate circumstances, allowing
efficient large-scale identification of X-ray source
optical counterparts. SDSS and RASS imaging catalog data
are autonomously cross-correlated, and software algorithms 
employed to automatically select, and assign priorities to, 
candidate optical counterparts for follow-on SDSS fiber 
spectroscopy. Application of this approach to
initial SDSS data has provided
homogeneous identification and RASS/SDSS flux and
spectroscopic data for a large sample of X-ray emitting quasars and other kinds of AGN. 
The combination of SDSS multicolor selection
and RASS data---and in some cases FIRST radio information---is 
highly ($>$75\%) efficient for selection of X-ray emitting quasars/AGN.
In an initial 1400~deg$^2$ of sky considered, more than 1200 plausible 
X-ray emitting quasars/AGN have been
optically identified, including numerous rare cases
such as 45 BL~Lac candidates and more than 130 NLS1s. As this initial area 
represents
only a fraction of the ultimate joint RASS/SDSS sky coverage,
$\sim10^4$ fully and homogeneously characterized 
X-ray source counterpart identifications 
may be anticipated to follow by completion of the SDSS survey.
The already large sample will allow for a variety of more detailed
studies of various AGN subclasses and individual objects of special
interest,
as well as for studies of ensemble correlations between optical and X-ray
wavebands. 

\bigskip
\bigskip
{\it Acknowledgments}:
Funding for the creation and distribution of the SDSS Archive has been
provided by the Alfred P. Sloan Foundation, the Participating Institutions, the
National Aeronautics and Space Administration, the National Science Foundation, 
the U.S. Department of Energy, the Japanese Monbukagakusho, and the Max Planck 
Society. The SDSS Web site is http://www.sdss.org/. 
The SDSS is managed by the Astrophysical Research Consortium (ARC) for
the Participating Institutions. The Participating Institutions
are The University of Chicago, Fermilab, the Institute for Advanced Study, the
Japan Participation Group, The Johns Hopkins University, Los Alamos National
Laboratory, the Max-Planck-Institute for Astronomy (MPIA), the Max-Planck-Institute for
Astrophysics (MPA), New Mexico State University, University of Pittsburgh,
Princeton University, the United States Naval Observatory, and the University
of Washington.

The authors thank the late Donald
Baldwin of the University of Washington for his tireless 
contributions to
the success of SDSS.


\newpage
\tiny
\begin{center}
\begin{table}[!htb]
\tablenum{1}
\caption{Observed Parameters of Broad-Line RASS/SDSS AGN\tablenotemark{a}}
\begin{tabular}{llrrrrrrrrrrrrr}
\tableline
\tableline
 RASS X-ray & SDSS optical & $u^*$ & $g^*$ & $r^*$ & $i^*$ & $z^*$ & opt & red- & X-ray & X & X & X & X & $f_x\times$ \\
 source & counterpart & & & & & & morph & shift & count & exp & HR1 & HR2 & like & $10^{13}$ \\
 RXS J & SDSS J& & & & & & & & rate & tim & & & & \\
 (1) & (2) & (3) & (4) & (5) & (6) & (7) & (8) & (9) & (10) & (11) & (12) & (13) & (14) & (15)\\
\tableline
000011.9+000223 & 000011.96+000225.2 & 17.83 & 17.58 & 17.69 & 17.65 & 17.66 & 6 & 0.479 & 0.0219 & 364 & -0.01 & -1.00 & 9 & 2.21 \\
000250.8+000824 & 000251.60+000800.5 & 19.98 & 18.95 & 18.11 & 17.64 & 17.48 & 3 & 0.107 & 0.0410 & 388 & -0.32 & 0.09 & 9 & 4.10 \\
000611.9$-$010648 & 000608.04$-$010700.8 & 18.95 & 18.54 & 18.44 & 18.43 & 18.55 & 6 & 0.948 & 0.0243 & 394 & -0.25 & -0.66 & 8 & 2.49 \\
000709.8+005328 & 000710.01+005329.0 & 17.18 & 16.87 & 16.64 & 16.69 & 16.05 & 6 & 0.316 & 0.1045 & 405 & 0.07 & 0.04 & 62 & 10.12 \\
000813.3$-$005752 & 000813.22$-$005753.3 & 18.97 & 18.29 & 17.66 & 17.08 & 16.89 & 3 & 0.139 & 0.0508 & 394 & -0.02 & 0.02 & 27 & 5.14 \\
\tableline
\end{tabular}
\tablenotetext{a}{[The complete version of this table is in the electronic
edition of the Journal. The printed edition contains only a sample.]}
\end{table}
\end{center}
\normalsize

\tiny
\begin{center}
\begin{table}[!htb]
\tablenum{2}
\caption{Derived Parameters of Broad-Line RASS/SDSS AGN\tablenotemark{a}}
\begin{tabular}{llrrrrrrrl}
\tableline
\tableline
 RASS X-ray & SDSS optical & $g^*_o$ & red- & $f_x^c \times$ & $log(L_x)$ & $log(l_{opt})$ & $log(l_{x})$ & $\alpha_{ox}$ & comment \\
 source &  counterpart & & shift & $10^{13}$ & & 2500\AA\ & 2~kev &  & \\
 RXS J & SDSS J & &  & & & & & \\
 (1) & (2) & (3) & (4) & (5) & (6) & (7) & (8) & (9) & (10)\\
\tableline
000011.9+000223 & 000011.96+000225.2 & 17.46 & 0.479 & 5.43 & 44.76 & 30.29 & 26.22 & 1.56 & LBQS-2357-0014\\
000250.8+000824 & 000251.60+000800.5 & 18.83 & 0.107 & 10.03 & 43.49 & 28.33 & 24.95 & 1.30 & ...\\
000611.9$-$010648 & 000608.04$-$010700.8 & 18.40 & 0.948 & 6.21 & 45.60 & 30.58 & 27.06 & 1.35 & radio\\
000709.8+005328 & 000710.01+005329.0 & 16.75 & 0.316 & 24.18 & 44.95 & 30.17 & 26.42 & 1.44 & radio,LBQS-0004+0036\\
000813.3$-$005752 & 000813.22$-$005753.3 & 18.17 & 0.139 & 12.70 & 43.84 & 28.84 & 25.31 & 1.36 & ...\\
\tableline
\end{tabular}
\tablenotetext{a}{[The complete version of this table is in the electronic
edition of the Journal. The printed edition contains only a sample.]}
\end{table}
\end{center}
\normalsize

\tiny
\begin{center}
\begin{table}[!htb]
\tablenum{3}
\caption{Observed Parameters of RASS/SDSS AGN Having Narrower
Permitted Emission\tablenotemark{a}}
\begin{tabular}{llrrrrrrrrrrrrr}
\tableline
\tableline
 RASS X-ray & SDSS optical & $u^*$ & $g^*$ & $r^*$ & $i^*$ & $z^*$ & opt & red- & X-ray & X & X & X & X & $f_x\times$ \\
 source & counterpart & & & & & & morph & shift & count & exp & HR1 & HR2 & like & $10^{13}$ \\
 RXS J & SDSS J& & & & & & & & rate & tim & & & & \\
 (1) & (2) & (3) & (4) & (5) & (6) & (7) & (8) & (9) & (10) & (11) & (12) & (13) & (14) & (15)\\
\tableline
002847.7+145142 & 002848.77+145216.3 & 19.66 & 18.88 & 18.44 & 17.98 & 17.71 & 3 & 0.089 & 0.0247 & 422 & 0.13 & -0.69 & 14 & 2.80 \\
003238.2$-$010056 & 003238.20$-$010035.2 & 18.90 & 18.25 & 17.84 & 17.41 & 17.22 & 3 & 0.092 & 0.0688 & 625 & -0.10 & 0.46 & 33 & 6.50 \\
003846.3+003430 & 003843.06+003451.7 & 19.82 & 18.43 & 17.95 & 17.67 & 17.32 & 3 & 0.043 & 0.0641 & 303 & -0.39 & 0.77 & 18 & 5.38 \\
004055.2+000039 & 004052.14+000057.2 & 18.37 & 18.20 & 18.04 & 18.11 & 17.88 & 6 & 0.405 & 0.0232 & 296 & -1.00 & 0.00 & 8 & 1.97 \\
004532.6$-$005807 & 004533.46$-$005808.8 & 18.78 & 18.56 & 18.32 & 17.97 & 18.04 & 6 & 0.138 & 0.0512 & 329 & -1.00 & 0.00 & 14 & 4.88 \\
\tableline
\end{tabular}
\tablenotetext{a}{[The complete version of this table is in the electronic
edition of the Journal. The printed edition contains only a sample.]}
\end{table}
\end{center}
\normalsize

\tiny
\begin{center}
\begin{table}[!htb]
\tablenum{4}
\caption{Derived Parameters of RASS/SDSS AGN Having Narrower
Permitted Emission\tablenotemark{a}}
\begin{tabular}{llrrrrrrrl}
\tableline
\tableline
 RASS X-ray & SDSS optical & $g^*_o$ & red- & $f_x^c \times$ & $log(L_x)$ & $log(l_{opt})$ & $log(l_{x})$ & $\alpha_{ox}$ & comment \\
 source &  counterpart & & shift & $10^{13}$ & & 2500\AA\ & 2~kev &  & \\
 RXS J & SDSS J & &  & & & & & \\
 (1) & (2) & (3) & (4) & (5) & (6) & (7) & (8) & (9) & (10)\\
\tableline
002847.7+145142 & 002848.77+145216.3 & 18.59 & 0.089 & 7.57 & 43.19 & 28.26 & 24.66 & 1.38 & radio,Sy1.5\\
003238.2$-$010056 & 003238.20$-$010035.2 & 18.17 & 0.092 & 15.24 & 43.53 & 28.46 & 24.99 & 1.33 & Sy1.5\\
003846.3+003430 & 003843.06+003451.7 & 18.36 & 0.043 & 11.64 & 42.71 & 27.70 & 24.17 & 1.36 & Sy2?\\
004055.2+000039 & 004052.14+000057.2 & 18.09 & 0.405 & 4.30 & 44.47 & 29.88 & 25.93 & 1.51 & NLS1\\
004532.6$-$005807 & 004533.46$-$005808.8 & 18.49 & 0.138 & 11.54 & 43.79 & 28.70 & 25.26 & 1.32 & ...\\
\tableline
\end{tabular}
\tablenotetext{a}{[The complete version of this table is in the electronic
edition of the Journal. The printed edition contains only a sample.]}
\end{table}
\end{center}
\normalsize

\tiny
\begin{center}
\begin{table}[!htb]
\tablenum{5}
\caption{Observed Parameters of RASS/SDSS BL~Lac Candidates\tablenotemark{a}}
\begin{tabular}{llrrrrrrrrrrrrr}
\tableline
\tableline
 RASS X-ray & SDSS optical & $u^*$ & $g^*$ & $r^*$ & $i^*$ & $z^*$ & opt & red- & X-ray & X & X & X & X & $f_x\times$ \\
 source & counterpart & & & & & & morph & shift & count & exp & HR1 & HR2 & like & $10^{13}$ \\
 RXS J & SDSS J& & & & & & & & rate & tim & & & & \\
 (1) & (2) & (3) & (4) & (5) & (6) & (7) & (8) & (9) & (10) & (11) & (12) & (13) & (14) & (15)\\
\tableline
003514.9+151513 & 003514.72+151504.1 & 17.32 & 16.93 & 16.58 & 16.28 & 16.04 & 6 & ... & 0.2358 & 359 & 0.35 & 0.25 & 235 & 27.08 \\
020106.3+003400 & 020106.17+003400.2 & 19.35 & 19.06 & 18.43 & 18.14 & 17.83 & 3 & 0.298 & 0.3390 & 400 & 0.07 & 0.04 & 312 & 31.21 \\
023812.9$-$092441 & 023813.67$-$092431.4 & 20.82 & 20.26 & 19.64 & 19.21 & 18.84 & 3 & 0.740 & 0.0345 & 224 & 0.72 & 0.42 & 17 & 3.32 \\
030433.0$-$005409 & 030433.95$-$005404.6 & 19.15 & 18.86 & 18.50 & 18.15 & 17.90 & 6 & 0.513 & 0.0559 & 201 & 0.57 & 0.80 & 21 & 7.22 \\
075436.4+391103 & 075437.07+391047.7 & 18.92 & 17.92 & 17.23 & 16.80 & 16.53 & 3 & 0.096 & 0.0344 & 422 & 0.06 & -0.68 & 18 & 3.96 \\
\tableline
\end{tabular}
\tablenotetext{a}{[The complete version of this table is in the electronic
edition of the Journal. The printed edition contains only a sample.]}\end{table}
\end{center}
\normalsize

\tiny
\begin{center}
\begin{table}[!htb]
\tablenum{6}
\caption{Derived Parameters of RASS/SDSS BL~Lac Candidates\tablenotemark{a}}
\begin{tabular}{llrrrrrrrl}
\tableline
\tableline
 RASS X-ray & SDSS optical & $g^*_o$ & red- & $f_x^c \times$ & $log(L_x)$ & $log(l_{opt})$ & $log(l_{x})$ & $\alpha_{ox}$ & comment \\
 source &  counterpart & & shift & $10^{13}$ & & 2500\AA\ & 2~kev &  & \\
 RXS J & SDSS J & &  & & & & & \\
 (1) & (2) & (3) & (4) & (5) & (6) & (7) & (8) & (9) & (10)\\
\tableline
003514.9+151513 & 003514.72+151504.1 & 16.66 & ... & 74.47 & ... & ...& ... & 1.27 & RBS-0082 \\
020106.3+003400 & 020106.17+003400.2 & 18.97 & 0.298 & 71.93 & 45.37 & 29.23& 26.83 & 0.92 & MS-0158.5+0019 \\
023812.9$-$092441 & 023813.67$-$092431.4 & 20.16 & 0.740 & 7.88 & 45.41 & 29.63& 26.87 & 1.06 & RX-J0238.2-0924,zunc \\
030433.0$-$005409 & 030433.95$-$005404.6 & 18.56 & 0.513 & 23.62 & 45.47 & 29.92& 26.93 & 1.15 & FBQS-J0304-0054,zunc \\
075436.4+391103 & 075437.07+391047.7 & 17.74 & 0.096 & 10.91 & 43.42 & 28.67& 24.89 & 1.45 & RX-J0754.6+3911 \\
\tableline
\end{tabular}
\tablenotetext{a}{[The complete version of this table is in the electronic
edition of the Journal. The printed edition contains only a sample.]}\end{table}
\end{center}

\normalsize

\begin{figure}[!htb]
\plotone{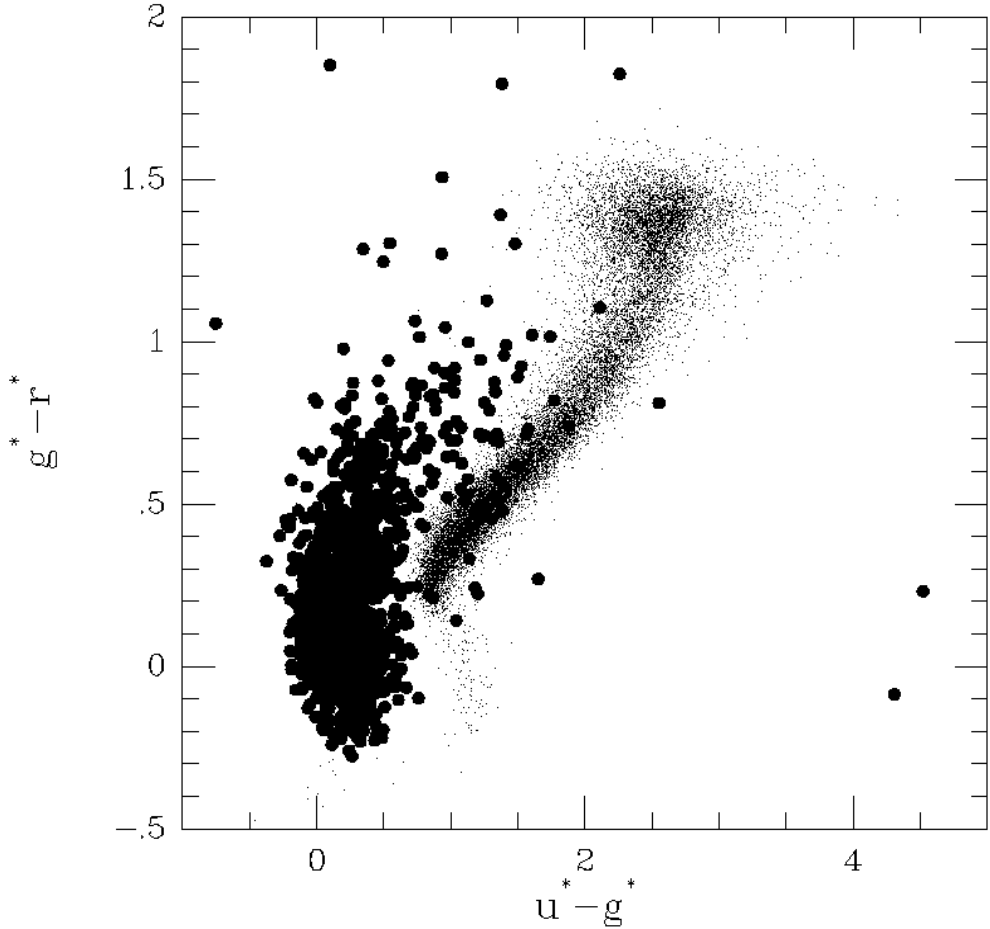}
\caption{ Odd colored SDSS optical objects (as well as coincidences with
FIRST radio sources) and positionally coincident with RASS X-ray 
positions are given high priority for SDSS spectroscopic follow-up.
The SDSS colors of 1225 spectroscopically-confirmed 
quasars and other X-ray emitting AGN in this initial sample are 
overplotted (large solid circles) for comparison on the locus 
of 10,000 anonymous SDSS stellar objects (small points). 
Note the clean color separation of the large majority of the confirmed 
quasar/AGN identifications. [For simplicity of display, three very red 
in $u-g$ objects (formally with PSF colors $u^*-g^*>5$) are not shown. These objects 
are extremely faint in $u$, so their precise $u-g$ colors are uncertain.]}

\end{figure}

\begin{figure}[!htb]
\plotone{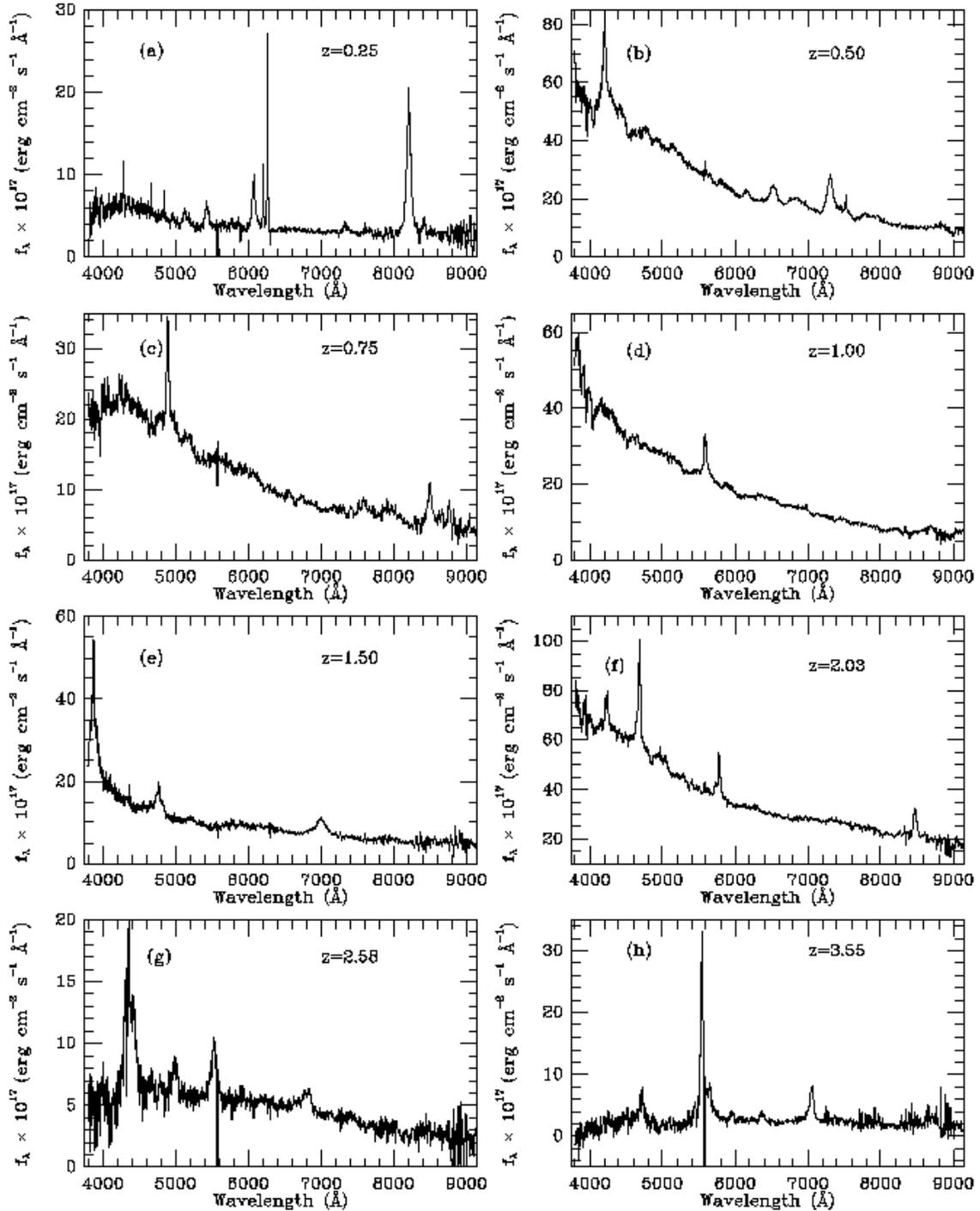}
\caption{Example SDSS optical spectra for confirmed
RASS/SDSS X-ray emitting quasars/AGN with (mainly)
broad permitted emission, as discussed in section 4.1.
Shown are spectra for typical objects covering a range of
redshift from $0.25$ to $3.5$. All 964 such objects cataloged/discussed
in section 4.1 have similar high-quality SDSS spectroscopy. 
(a) SDSS J233024.73+011602.3; (b) SDSS J234440.03$-$003231.6;
(c) SDSS J103245.16+644154.2; (d) SDSS J170035.42+632522.6;
(e) SDSS J172424.36+571531.0; (f) SDSS J150837.11+604955.8;
(g) SDSS J155104.23+521637.7; (h) SDSS J125227.28+001001.9.}
\end{figure}

\begin{figure}[!htb]
\resizebox{.75\textwidth}{!}{\rotatebox{-90}{\plotone{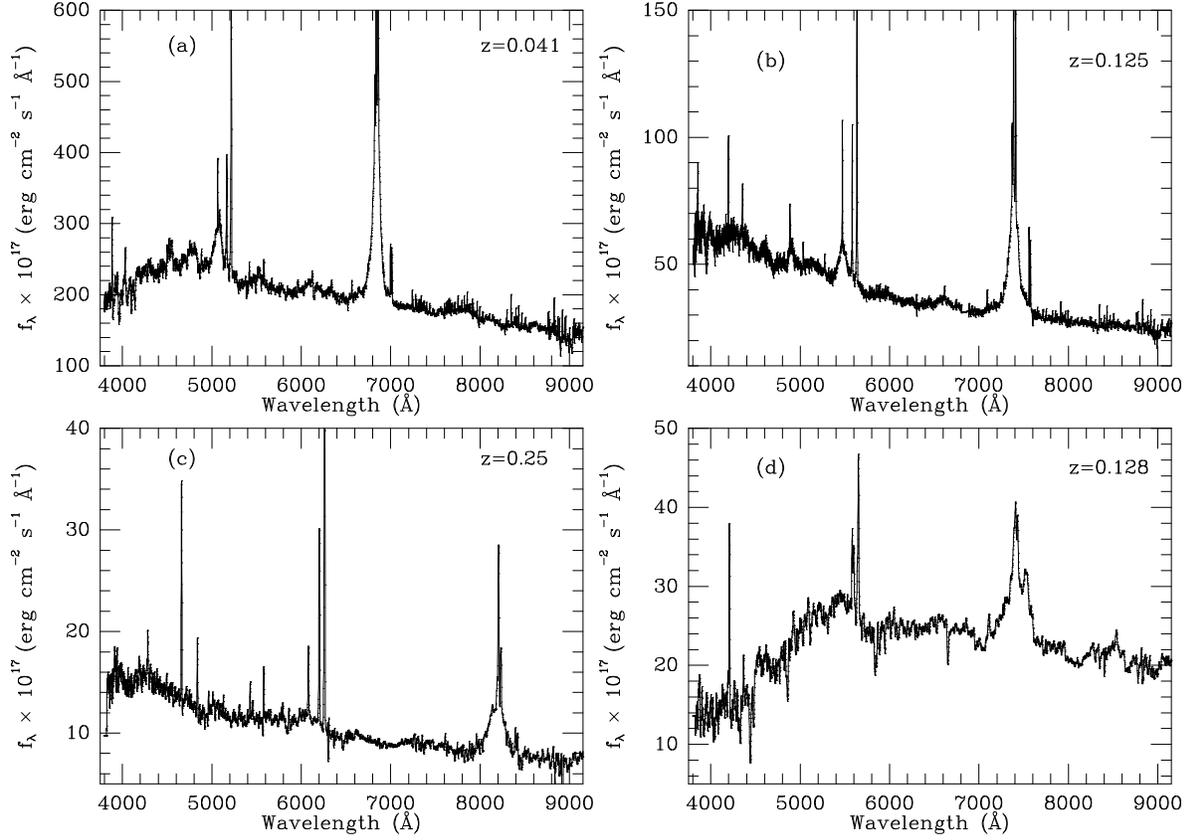}}}
\caption{ Example SDSS optical spectra for confirmed
RASS/SDSS X-ray emitting quasars/AGN with both broad and narrower
permitted emission components, chosen from 
the 216 objects cataloged/discussed in section 4.2.
Shown are selected spectra for intermediate 
Seyfert types between Sy~1.5 and 1.9.
(a) SDSS J101912.57+635802.7; (b) SDSS J112841.60+633551.3;
(c)~SDSS J130550.51$-$012331.6; (d) SDSS J082133.60+470237.2.}
\end{figure}

\begin{figure}[!htb]
\resizebox{.75\textwidth}{!}{\rotatebox{-90}{\plotone{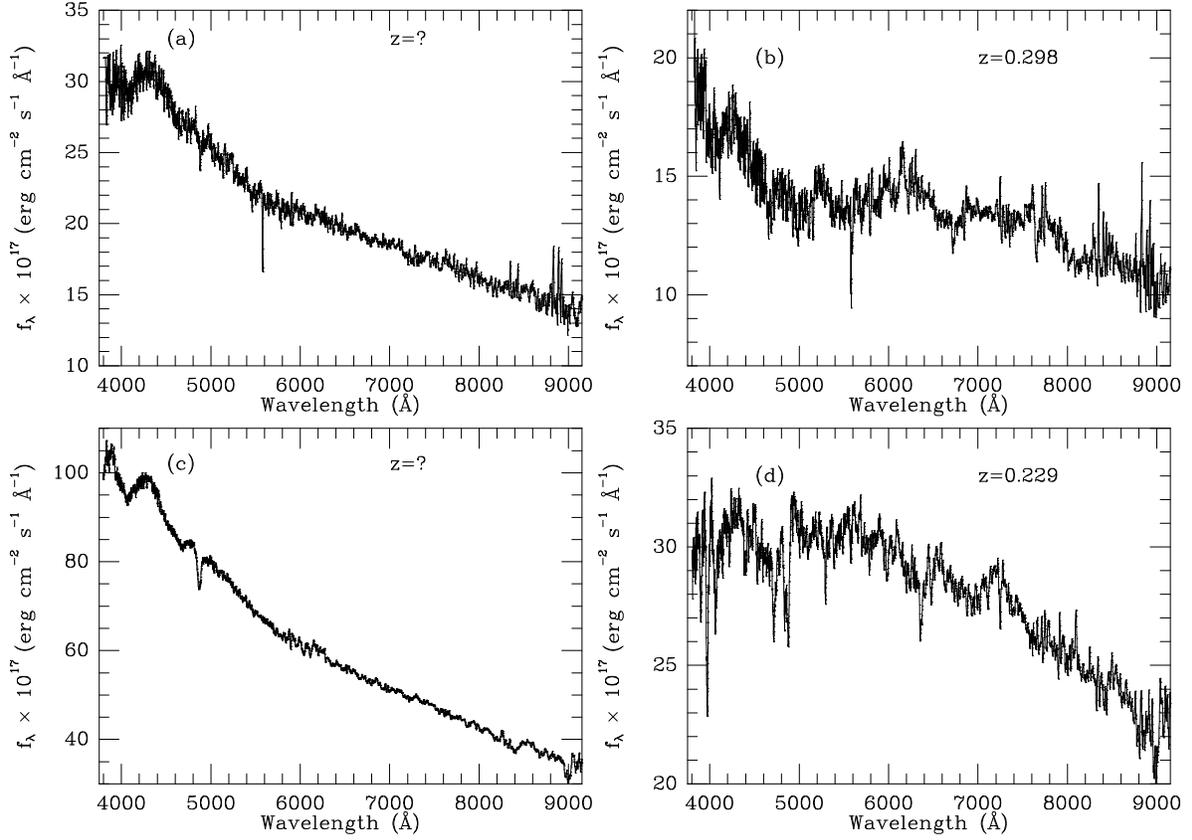}}}
\caption{Example SDSS optical spectra of
objects included in the RASS/SDSS sample of X-ray emitting 
BL~Lacs and candidates; see text
for discussion of selection. A total of 45 such candidates
are presented in this initial RASS/SDSS catalog.
(The apparent broad feature near 4300 \AA\ is 
due to a systematic error in fluxing of the spectra).
(a) SDSS J131106.47+003510.0; (b) SDSS J020106.17+003400.2;
(c) SDSS J003514.72+15`1504.1; (d) SDSS J094542.23+575747.7.}
\end{figure}

\begin{figure}[!htb]
\resizebox{.75\textwidth}{!}{\rotatebox{-90}{\plotone{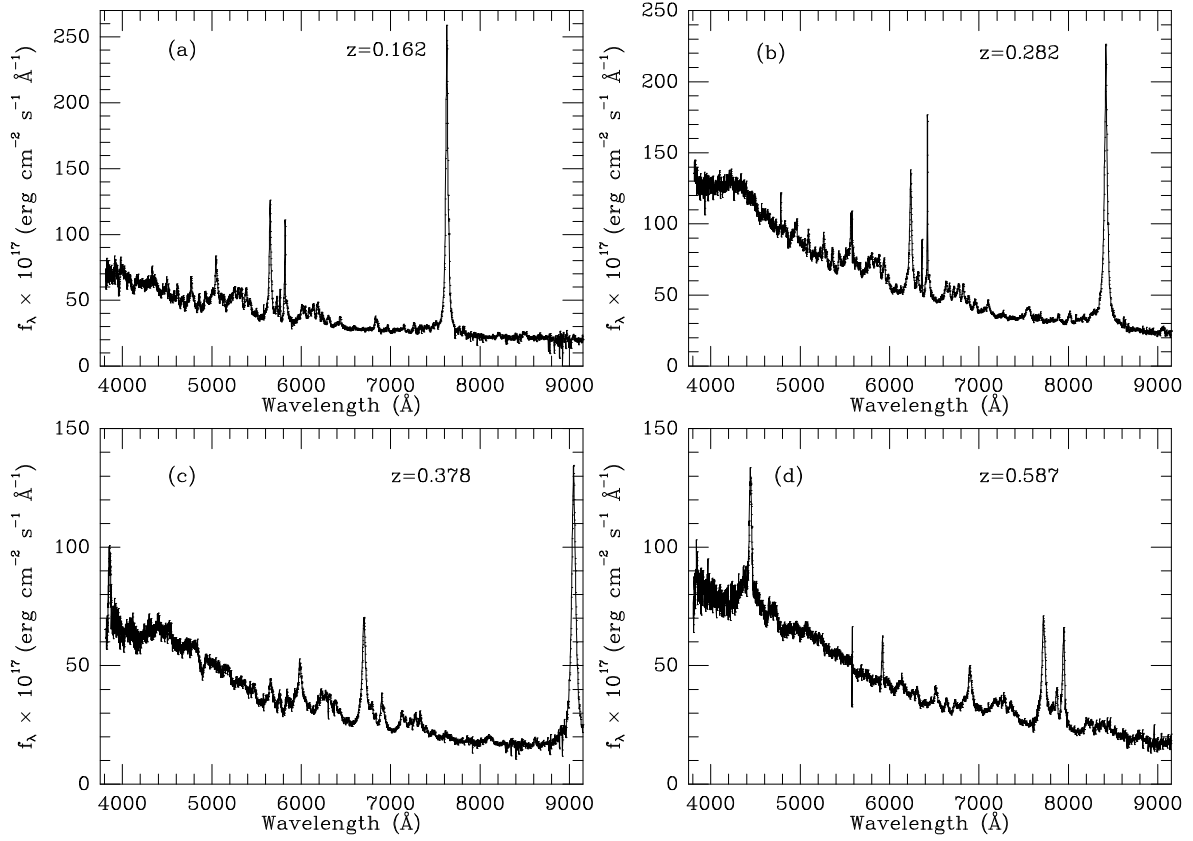}}}
\caption{Example SDSS optical spectra for
RASS/SDSS X-ray emitting NLS1s. There are 133 such confident NLS1s
(and another 36 possible ones) in this initial RASS/SDSS sample.
(a) SDSS J090137.99+532051.1; (b) SDSS J125100.45+660326.7;
(c) SDSS J024250.85$-$075914.3; (d) SDSS J134948.39$-$010621.8.}
\end{figure}

\begin{center}
\leavevmode
\begin{figure}[!htb]
\vspace*{-0.5cm}
\resizebox{.7\textheight}{!}{\rotatebox{0}{\plotone{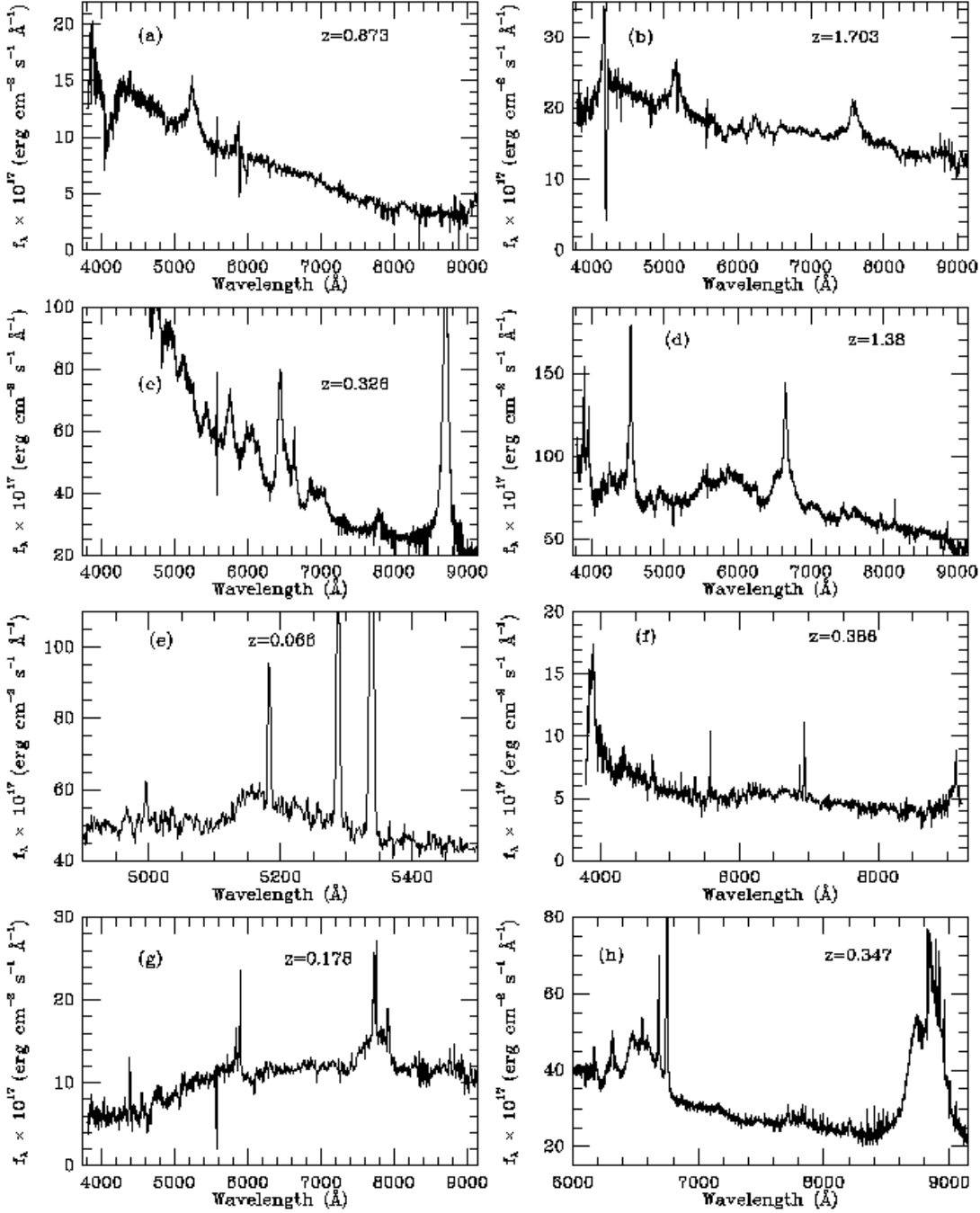}}}
\vspace*{-0.5cm}
\caption{Selected other X-ray emitting AGN with unusual 
optical SDSS spectra.
(a) SDSS J171216.28+660211.7 and (b) SDSS J150935.97+574300.5
are examples of {\it possible} X-ray emitting quasars with 
BALs or mini-BALs, though certainly not yet confirmed
definitively. (c) SDSS J134251.60$-$005345.2 and 
(d) SDSS J091301.01+525928.9 are examples of strong Fe quasars, with
permitted line widths in excess of those usually associated with
NLS1s. (e)~SDSS J172533.07+571645.5 is an example AGN with an asymmetric 
H$\beta$ line profile. (f) SDSS J113615.08$-$002314.3 is an AGN with 
broad weak H$\beta$ characteristic of Sy~1.8, but which has both strong and
broad H$\alpha$ and MgII. (g) SDSS J155350.44+562230.8 is an
example object with very wide line profiles of order 20000~km/sec.
(h) SDSS J075407.95+431610.5 is an example of an X-ray emitting
AGN in the sample with
highly unusual multiple-peaked emission lines profiles. }
\end{figure}

\begin{figure}[!htb]
\leavevmode
\vskip 1.4in
\plotone{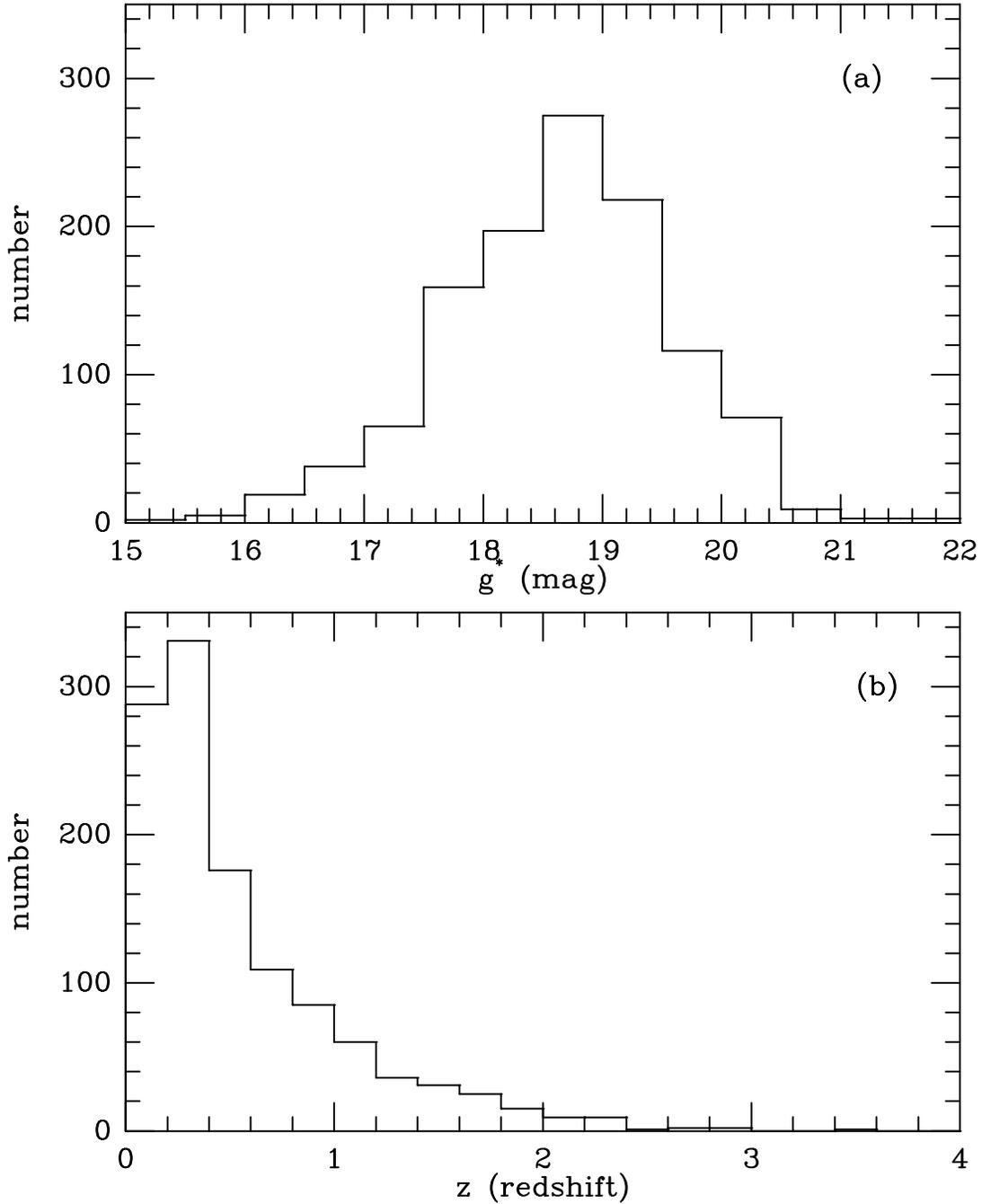}
\caption{The distributions of (a) SDSS $g^*$-band magnitudes, and
(b) redshifts, for 1180 optical counterparts of RASS/SDSS X-ray emitting AGN
(BL~Lac candidates excluded here). 
Redshifts are obtained
from the high-quality follow-up SDSS optical spectroscopy.
The median magnitude and redshift are typical of similar
identification surveys having comparable X-ray depths. However, the
large sample size permits inclusion of a significant number of higher 
redshift objects as well.
 }
\end{figure}

\begin{figure}[!htb]
\leavevmode
\vskip 1.4in
\plotone{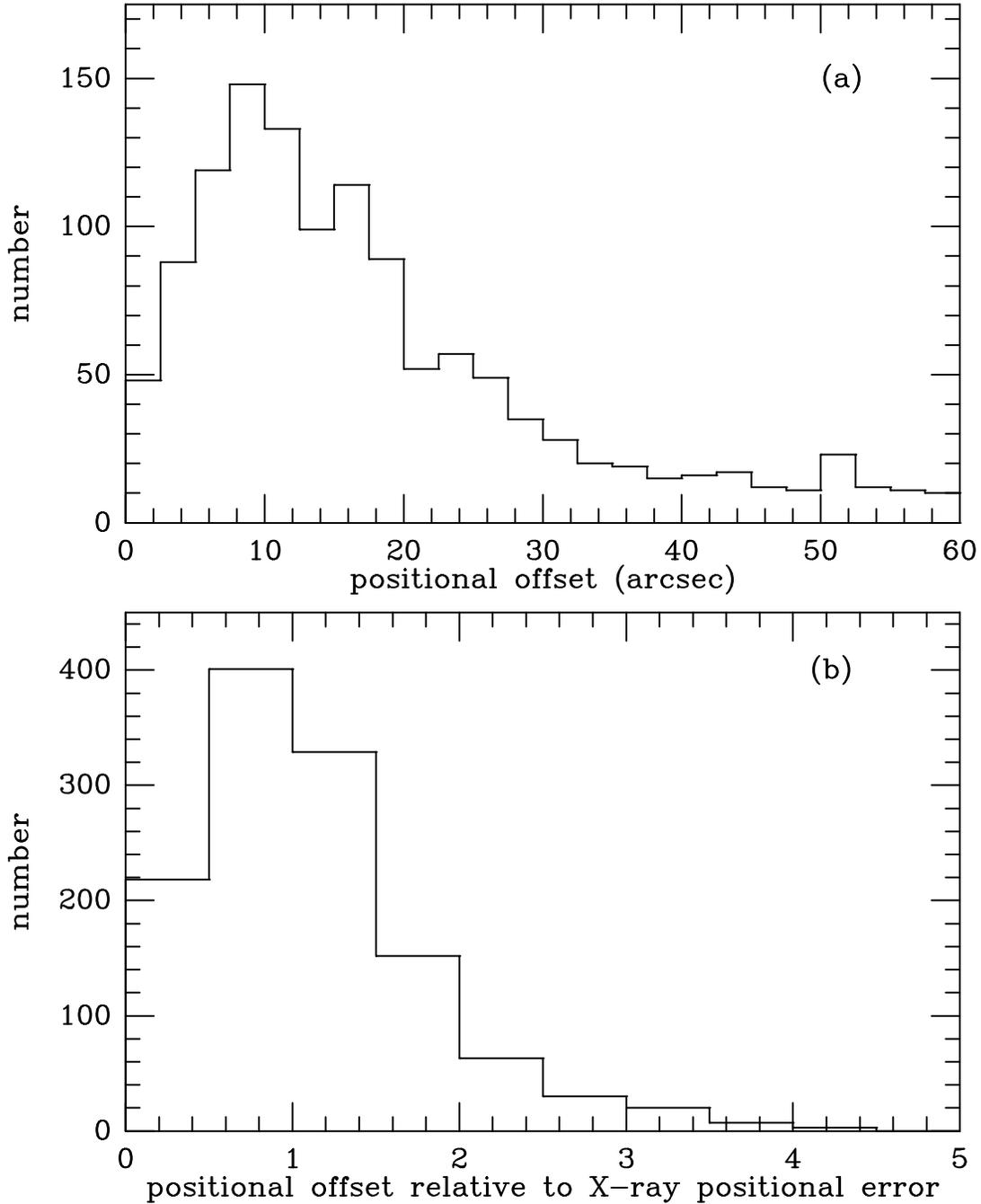}
\caption{The distribution of positional differences between the
SDSS optical positions of the 1225 quasar/AGN/BL~Lac suggested identifications are
approximately as expected if these are (statistically) the proper
identifications. (a) The positional offsets in arcseconds
are approximately as expected for the RASS positional accuracy, accounting
for the range of X-ray detection likelihoods included. (b) The distribution 
of relative positional differences between the
SDSS optical and RASS X-ray positions of the quasar/AGN/BL~Lac identifications, in this
case normalized to the expected RASS X-ray source positional error.
(For ease of display, 2 objects with relative offsets $>$5 are excluded 
from this plot). }
\end{figure}

\begin{figure}[!htb]
\leavevmode
\vskip 1.0in
\resizebox{.6\textheight}{!}{\rotatebox{0}{\plotone{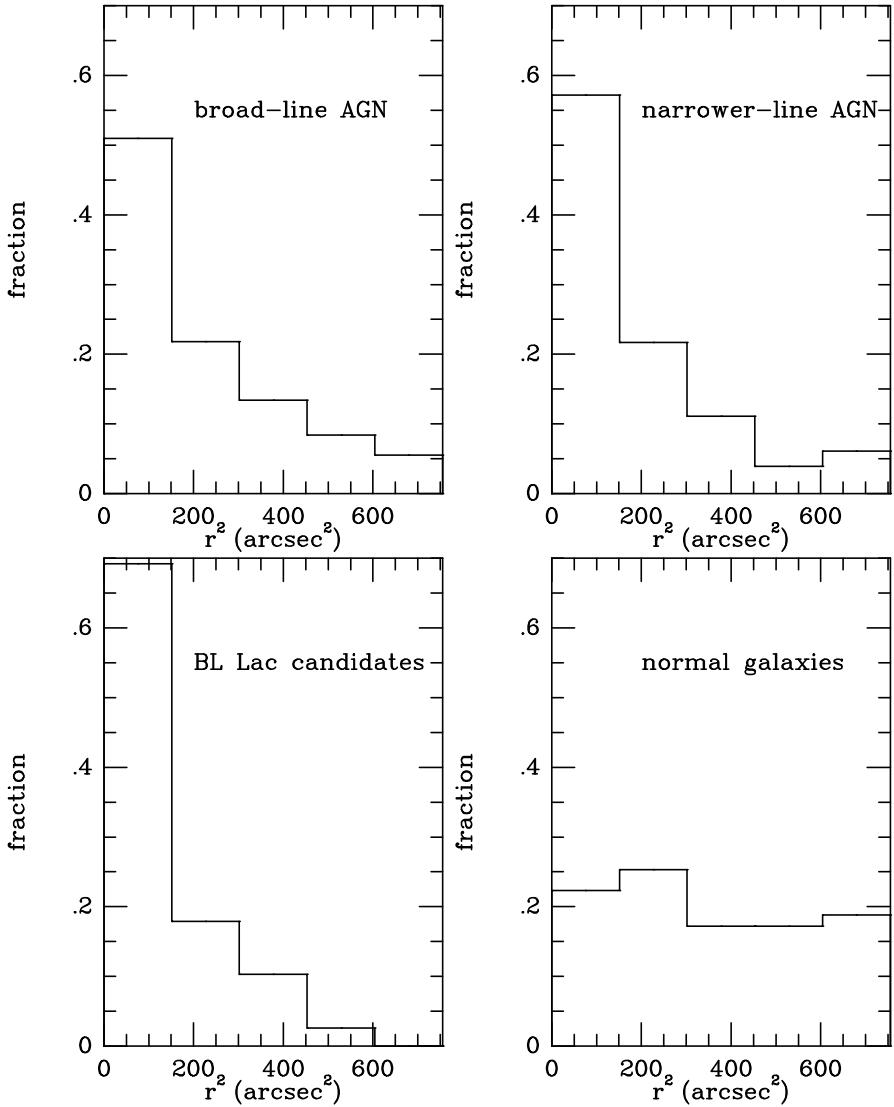}}}
\caption{Another measure of the distribution of positional 
offsets between RASS X-ray sources and various extragalactic 
SDSS optical objects (having SDSS spectra). Plotted are
histograms of the distributions of the squares, $r^2$, of the positional 
offsets between the SDSS optical positions and the RASS X-ray source positions; 
we count the fraction of objects falling within equal area 
annuli 
offset from the RASS X-ray source positions.
(We conservatively limit consideration to objects within $r<27.5''$ 
where all relevant algorithms for SDSS spectroscopy may select targets,
as discussed in section 6). Separate $r^2$ histograms are
shown for quasars/AGN with predominant broad-lines discussed 
in section 4.1 (upper left), quasars/AGN with narrower permitted 
lines discussed in section 4.2 (upper right), as well as BL~Lac 
candidates discussed in section 4.3 (lower left); in each of these
cases, the
histograms for the AGN are very strongly-peaked at small $r^2$ values, 
as expected if these AGN are statistically the proper (i.e., with low 
contamination) X-ray source identifications. For comparison, the lower 
right panel shows the histogram for normal SDSS galaxies 
(which, as anticipated, shows at most only a very weak statistical 
correlation with RASS X-ray sources). }
\end{figure}

\begin{figure}[!htb]
\leavevmode
\vskip .5in
\resizebox{.6\textheight}{!}{\rotatebox{0}{\plotone{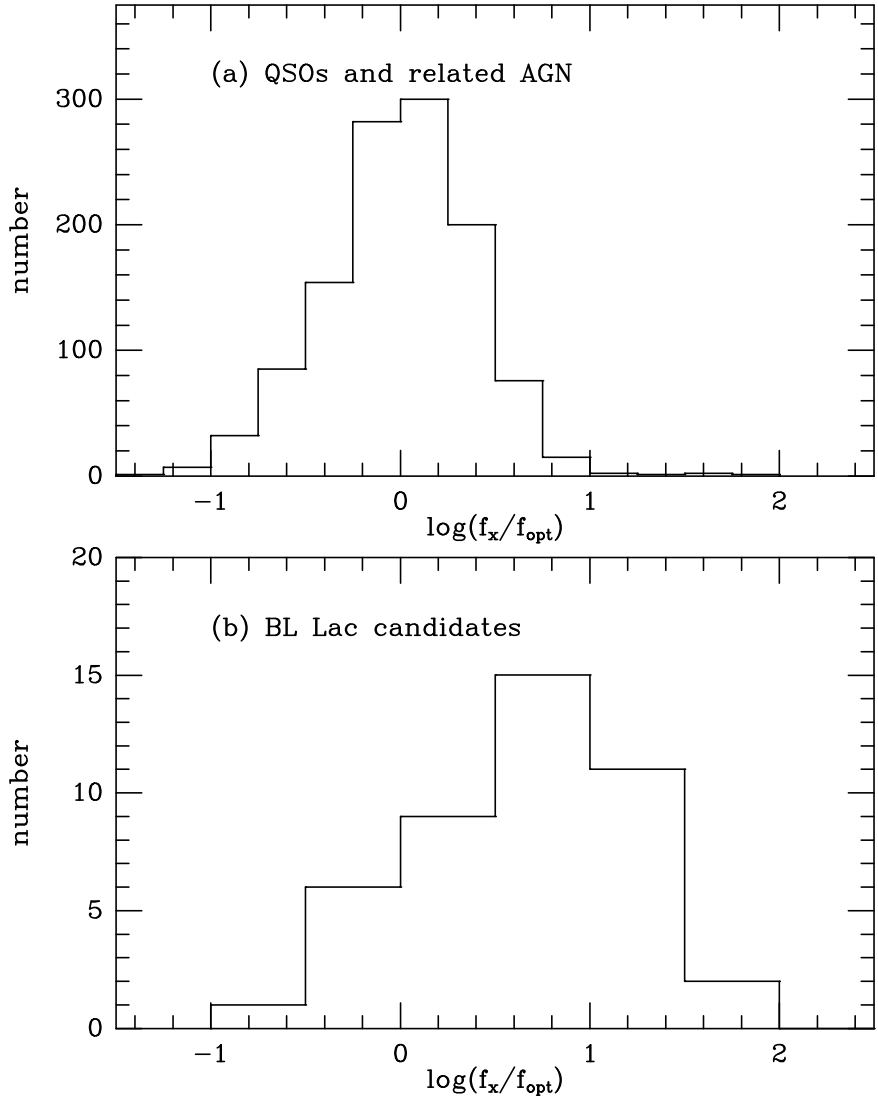}}}
\caption{(a) The distribution of $f_{x}/f_{opt}$ 
for the 1158 suggested quasar/AGN identifications having observed broad
emission line regions (Sy~2 candidates are excluded) 
discussed in sections 4.1
and 4.2. As expected if these are the proper identifications,
typical quasars/AGN are found to emit approximately as much energy in the X-ray as in the
optical band. (b) The distribution of $f_{x}/f_{opt}$ for the 
RASS/SDSS BL~Lac candidate identifications discussed in section 4.3. The 
distribution is
similar to that found in other X-ray selected BL~Lac surveys
(though markedly different than that of the emission-line
quasars/AGN shown Figure~9a). }
\end{figure}

\begin{figure}[!htb]
\vskip 1.0in
\resizebox{.6\textheight}{!}{\rotatebox{0}{\plotone{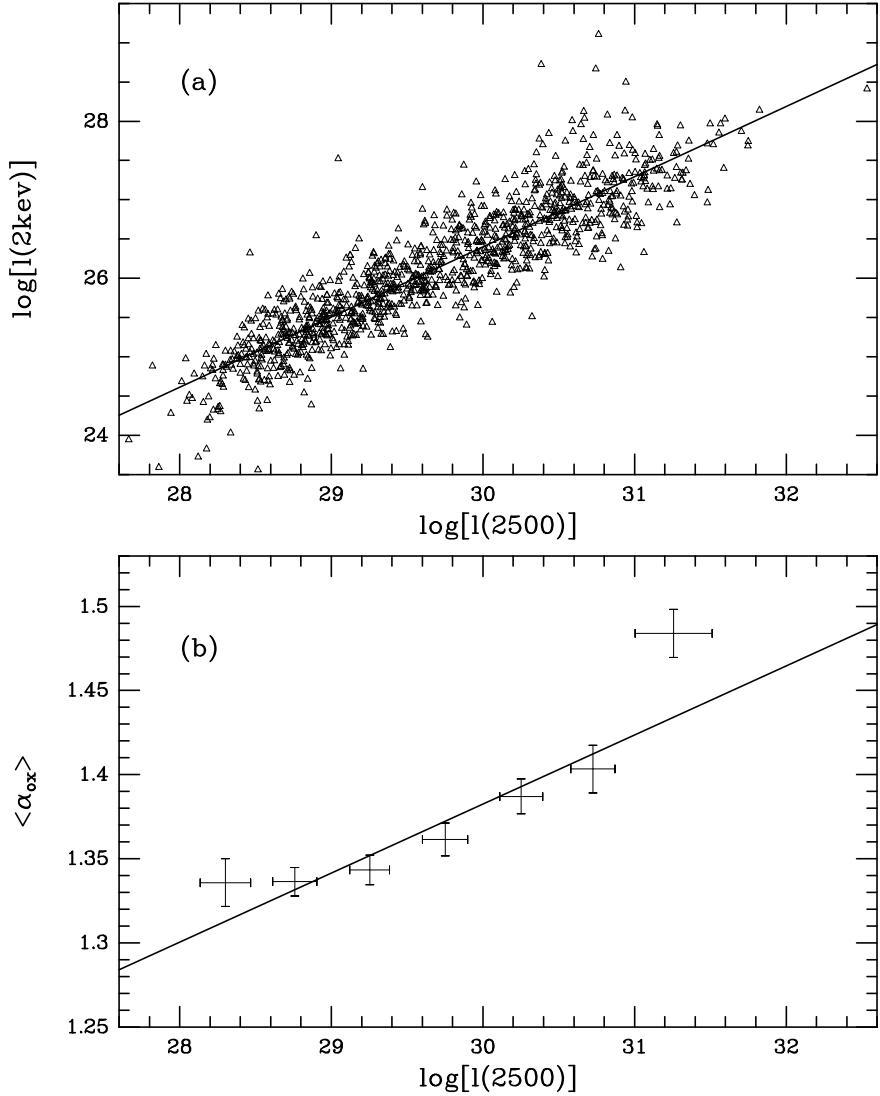}}}
\caption{ The long-recognized non-linear relationship between X-ray
and optical wavebands
(logarithms of monochromatic luminosities in cgs units at 2~kev and 2500\AA\
are shown) is also seen here in our large X-ray selected
sample of 1158 quasars/AGN (with broad emission line regions; Sy~2 and BL~Lac 
candidates are excluded). (a) The solid line is a 
least-squares fit to the $l_{2kev}$ versus  $l_{2500}$ data, with slope 
$0.89\pm0.01$. (b) Similar consistent results are obtained when 
alternately regressing $\alpha_{ox}$ against optical luminosity, and in
this case  linear regression yields  $\alpha_{ox} \propto l_{opt}^{0.041\pm0.005}$, 
equivalent as expected to $l_{x} \propto l_{opt}^{0.89}$.
The solid line is the  best fit regression relation (fit to all
1158 points), while the error bars show the mean and the standard error
in the
mean value of $<\alpha_{ox}>$, as well as the mean and standard deviations in optical
luminosity, when considering averages taken in various optical luminosity
bins (typically with 70 to 220 points per bin). There is an indication of
a possibly more complex relation than assumed in the simple regression
model. }
\end{figure}
\end{center}

\end{document}